\documentclass[intlimits,twoside,a4paper]{article}

\usepackage[cp1251]{inputenc}
\usepackage{xcolor}
\usepackage[eqsecnum]{cmpj3}

\usepackage{bm}

\issue{2020}{23}{4}{43709}
\doinumber{10.5488/CMP.23.43709}
\title[DMRG study of exciton condensation in the extended Falicov-Kimball
model]{DMRG study of exciton condensation in the extended Falicov-Kimball
model}
\author[P. Farka\v sovsk\'y]{P. Farka\v sovsk\'y}
\address{Institute  of  Experimental  Physics,  Slovak   Academy   of
Sciences,
Watsonova 47, 043 53 Ko\v {s}ice, Slovakia}
\date{Received May 15, 2020, in final form August 13, 2020}
\begin{document}

\maketitle

\begin{abstract}
The formation and condensation of excitonic bound states of conduction-band
electrons and valence-band holes surely belongs to one of the most
exciting ideas of contemporary solid state physics. In this short review 
we present the latest progress in this field reached by the 
density-matrix-renormalization-group (DMRG) calculations within 
various extensions of the Falicov-Kimball model.
Particular attention is paid to a description of crucial mechanisms
(interactions) that affect the stability of the excitonic phase,
and namely: (i) the interband $d$-$f$ Coulomb interaction, 
(ii) the $f$-electron hopping, (iii) the nonlocal hybridization 
with odd and even parity, (iv) combined effects of the local and nonlocal 
hybridization, (v) the nearest-neighbor Coulomb interaction between 
$d$ and $f$ electrons and (vi) the correlated hopping.
The relevance of numerical results obtained within different
extensions of the Falicov-Kimball model for a description of 
the real $d$-$f$ materials is widely discussed.

\end{abstract}
%\thanks{PACS nrs.: 71.27.+a, 71.28.+d, 71.30.+h}

\section{Introduction}

The formation of excitonic quantum condensates is an intensively
studied continuous problem in condensed matter physics~\cite{Blatt,Keldysh,
Mos,Litt}. Whilst theoretically predicted a long time ago~\cite{theo}, 
no conclusive experimental proof of the existence of the excitonic 
condensation has been achieved yet. However, the latest 
experimental studies of materials with strong electronic correlations 
showed that promising candidates for the experimental 
verification of the excitonic condensation could be 
TmSe$_{0.45}$Te$_{0.55}$~\cite{Wachter, Wachter1}, $1T$-TiSe$_2$~\cite{Monney, Monney1,Monney2,Monney3},
Ta$_2$NiSe$_5$~\cite{Wakisaka}, or a double bilayer graphene system~\cite{Perali}. In this regard, the mixed valence compound
TmSe$_{0.45}$Te$_{0.55}$ was argued to exhibit a pressure-induced 
excitonic instability, related to an anomalous increase in the 
electrical resistivity~\cite{Wachter, Wachter1}. In particular,  detailed studies 
of the pressure-induced semiconductor-semimetal transition in this 
material [based on the Hall effect, electrical and thermal (transport)
measurements] showed that excitons are created in a large quantity and condense 
below 20~K. On the other hand, in the layered transition-metal dichalcogenide 
$1T$-TiSe$_2$, a BCS-like electron-hole pairing was considered as the driving 
force for the periodic lattice distorsion~\cite{Monney,Monney1,Monney2,Monney3}. 
Moreover, quite recently, the excitonic-insulator state was probed 
by angle-resolved photoelectron spectroscopy in the semiconducting 
Ta$_2$NiSe$_5$ compound~\cite{Wakisaka}. These results have stimulated 
further experimental and theoretical studies with regard to the formation 
and possible condensation of excitonic bound states of electron and holes
in correlated systems. At present, it is generally accepted 
that the minimal theoretical model for a description of excitonic 
correlations in these materials could be the Falicov-Kimball
model~\cite{Falicov} and its extensions which were successfully used 
 in the past years to test the exciting idea of electronic 
ferroelectricity~\cite{P1,P2,Cz,F1,F2,Zl,Fr,B1,B2,F3,Schneider}
that is directly related with the formation of an excitonic 
insulator~\cite{Z1,Phan,Seki,Z2,Kaneko1,Kaneko2,Ejima,Kopec,Kunes,Golosov}. 
In its original form, the Falicov-Kimball model describes a two-band system of
localized $f$ electrons and itinerant $d$ electrons with short-ranged
$f$-$d$ Coulomb interaction $U$:
\begin{equation}
H_0=\sum_{ij}t_{ij}d^+_id_j+U\sum_if^+_if_id^+_id_i+E_f\sum_if^+_if_i\,,
\label{eq1}
\end{equation}
where $f^+_i$, $f_i$ are the creation and annihilation
operators  for an electron in  the localized state at 
lattice site $i$ with the binding energy $E_f$ and $d^+_i$,
$d_i$ are the creation and annihilation operators
of the itinerant spinless electrons in the $d$-band
Wannier state at site $i$. 

The first term of (\ref{eq1}) is the kinetic energy 
corresponding to quantum-mechanical hopping of the itinerant $d$ electrons   
between sites $i$ and $j$. These intersite hopping
transitions are described by the matrix  elements $t_{ij}$,
which are $-t_d$ if $i$ and $j$ are the nearest neighbors and
zero otherwise (in what follows all parameters are measured
in units of $t_d$). The second term represents the on-site
Coulomb interaction between the $d$-band electrons with density
$n_d=\frac{1}{L}\sum_id^+_id_i$ and the localized
$f$ electrons with density $n_f=\frac{1}{L}\sum_if^+_if_i$,
where $L$ is the number of lattice sites. The third  term stands
for the localized $f$ electrons whose sharp energy level is $E_f$.

Since in this simple model, the 
local occupation number $f^+_if_i$ commutates with the total 
Hamiltonian of the system, the local $f$-electron number is a strictly
conserved quantity and thus the $d$-$f$ electron coherence cannot be
established in such a system. If hybridization 
$H_V=V\sum_id^+_if_i+f^+_id_i$ between both bands is
included, the $f$ charge occupation is no longer a good quantum number,
and it is possible to build coherence between $d$ and $f$ electrons.   
Hybridization between the itinerant $d$ and localized $f$ states, however,
is not the only way to develop $d$-$f$ coherence. Theoretical works
of Batista et al.~\cite{B1,B2} showed that the ground state with 
a spontaneous electric polarization can also be induced by the
nearest-neighbor $f$-electron hopping $H_{t_f}=-t_f\sum_{<i,j>}f^+_if_j$,
but only for dimensions $D>1$.
In the strong coupling limit, this result was proven by mapping the
extended Falicov-Kimball model into the $xxz$ spin 1/2 model with a magnetic field along 
the $z$-direction, while in the intermediate coupling regime the
ferroelectric state was identified numerically by constrained path
Monte Carlo (CPMC) technique. 
Based on  these results, the authors postulated
the following conditions that favour the formation of the electronically
driven ferroelectric state: (a) The system must be in a mixed-valence regime
and the two bands involved must have different parity. (b) It is best, 
though not necessary, if both bands have similar bandwidths. (c) A local
Coulomb repulsion ($U$) between the different orbitals is required. 

Later on this model was  extensively used to describe different phases
in the ground state and special properties of the excitonic  
phase~\cite{Z1,Phan,Seki,Z2,Kaneko1,Kaneko2,Ejima}. 
It was found that the ground state phase diagram 
exhibits a very simple structure consisting of only four phases, and   
namely, the full $d$ and $f$ band insulator (BI), the excitonic
insulator (EI), the charge-density-wave (CDW) and the  staggered
orbital order (SOO). The EI is characterized by a nonvanishing 
$\langle d^+f \rangle$ average. The CDW is described by a periodic 
modulation in the total electron density of both $f$ and $d$ electrons, 
and the SOO is characterized by a periodic modulation in the
difference between the $f$ and $d$ electron densities.

In this article we focus our attention on the properties of 
the EI phase induced by local hybridization $V$ in the one dimension. 
Although it is generally known that there is no nonvanishing 
$P_{df}=\langle d^+f \rangle $ expectation value in the limit of vanishing 
hybridization (no spontaneous hybridization), the studies that we  
performed in the past years on various extensions of the original 
Falicov-Kimball model showed that it is possible to dramatically enhance  
excitonic correlations in the limit of small, but finite $V$ by additional 
interactions/factors~\cite{Farky_epl,Farky_prb,Farky_ssc,Farky_epjb}.
The effects of most important interactions are discussed in this
review. In particular, there are: (i) the interband $d$-$f$ Coulomb
interaction, (ii) the $f$-electron hopping, (iii) the nonlocal
hybridization with odd and even parity, (iv) combined effects of 
the local and nonlocal hybridization, (v) the nearest-neighbor Coulomb 
interaction between $d$ and $f$ electrons and (vi) the correlated hopping.     
The main goal of this review is not to examine the possibilities of spontaneous
symmetry breaking (a spontaneous hybridization) in various extensions
of the Falicov-Kimball model, but to show how these extensions (different 
interaction terms) influence the properties of the excitonic phase
induced by local hybridization.
All presented results were obtained within the 
density-matrix-renormalization-group (DMRG) method. where we typically 
keep up to 500 states per block, although in the numerically more difficult 
cases (where the DMRG results converge slower), we keep up to 1000 states. 
Truncation errors~\cite{White}, given by the sum of the density
matrix eigenvalues of the discarded states, vary from $10^{-6}$ in the worse 
cases to zero in the best cases.
\newpage
%\vspace{-0.3cm}
\section{Results and Discussion}
\subsection{Effects of interband Coulomb interaction}
\begin{figure}[!t]
\begin{center}
\includegraphics[width=5.8cm]{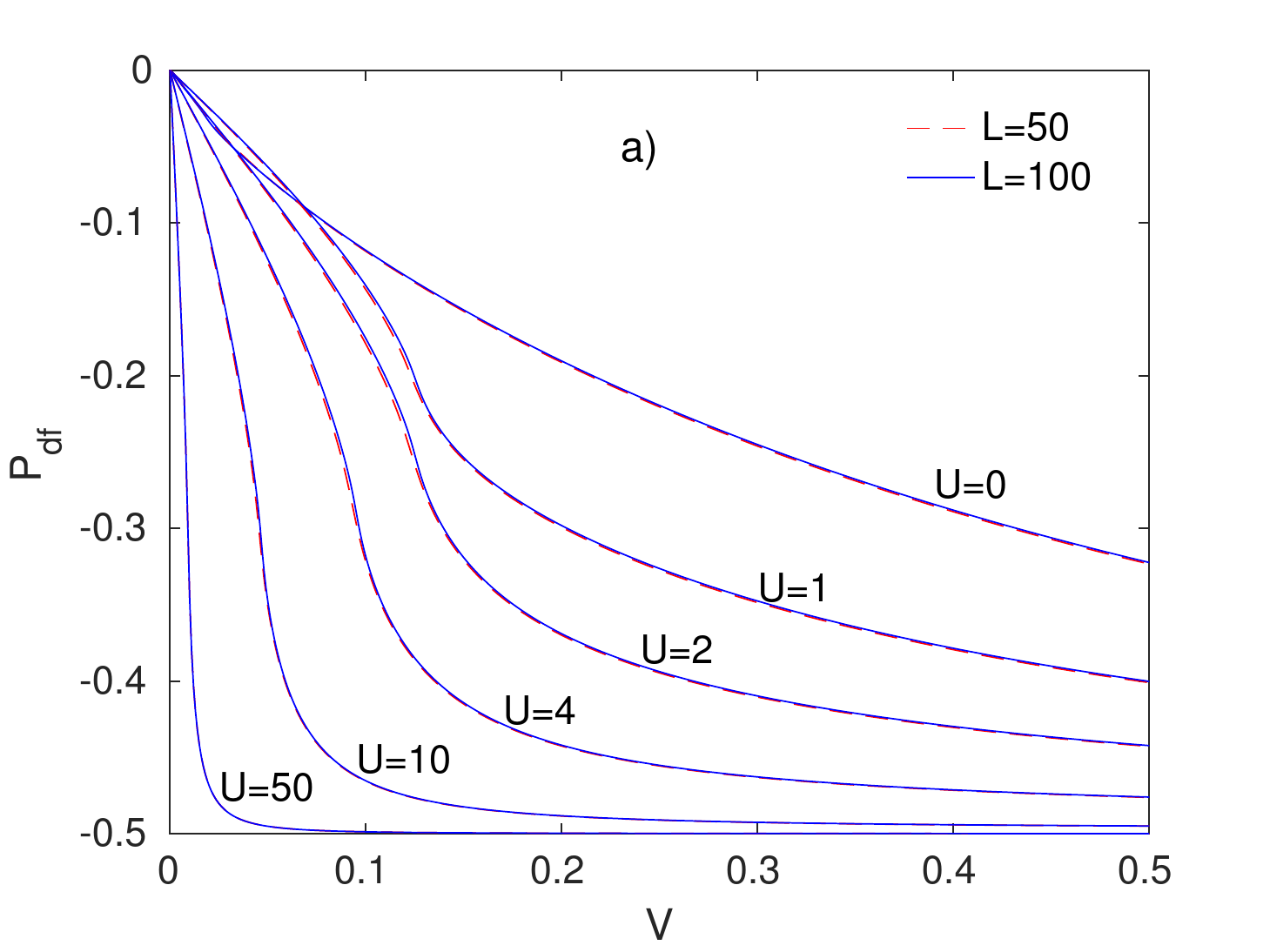}
\includegraphics[width=5.8cm]{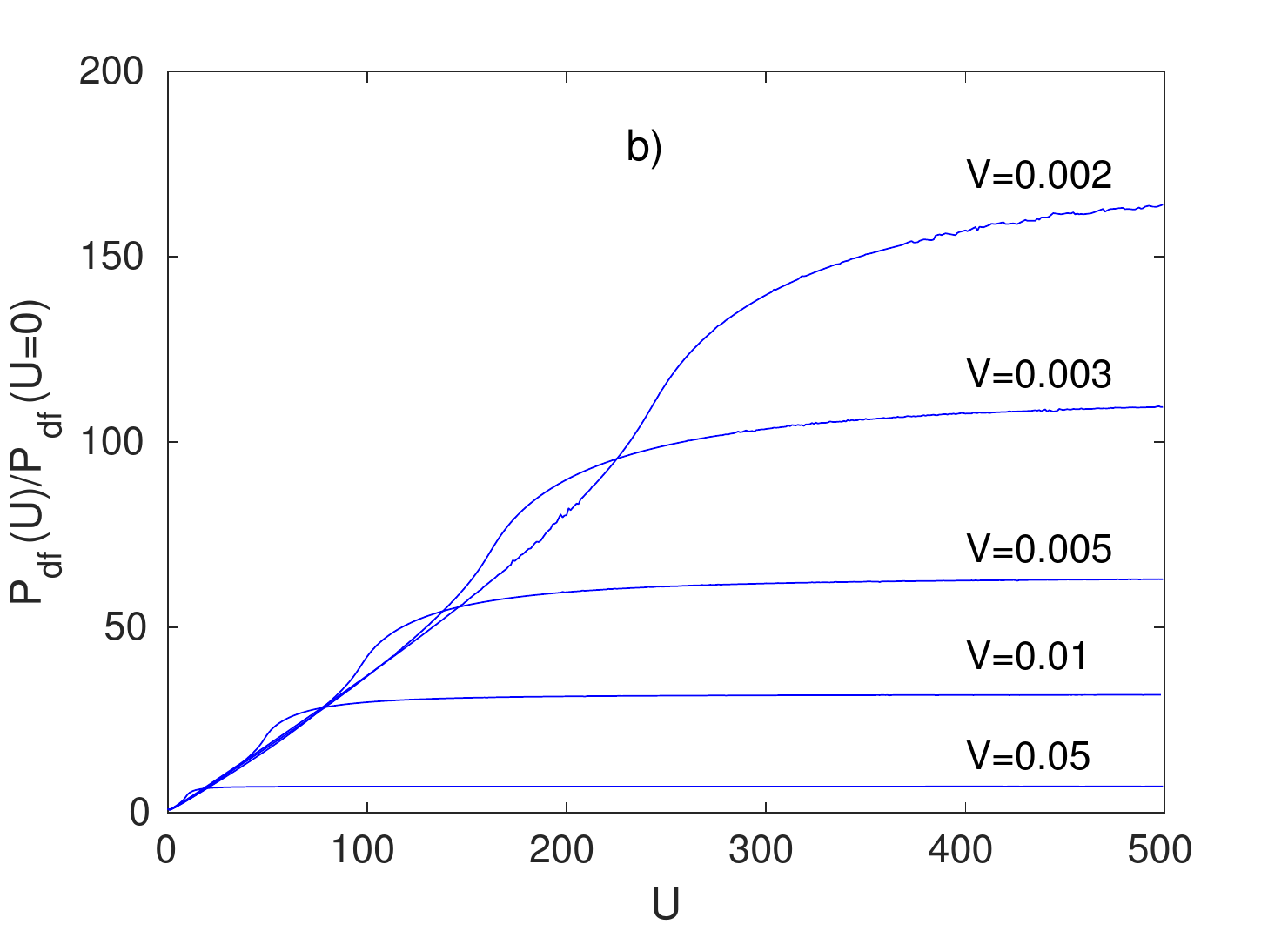}
\end{center}
\vspace*{-0.6cm}
\caption{\small (Colour online) a) The hybridization dependence of the $d$-$f$-excitonic average
$P_{df}=\langle d^+_if_i \rangle$ in the extended Falicov-Kimball model calculated for six
different values of $U$ and two different values of~$L$. The symmetric case $E_f=0$.
b) The ratio of the interacting $P_{df}(U)$ and non-interacting 
$P_{df}(U=0)$ excitonic average as a function of $U$ calculated for 
several selected values of local hybridization $V$ on the cluster of $L=100$
sites~\cite{Farky_epl}.}
\label{fig1}
\end{figure}

Let us start our review with the discussion of effects of the Coulomb
interactions~\cite{Farky_epl}. In this case, Hamiltonian consists of two terms: $H_0$,
which is given by (\ref{eq1}) and $H_V=V\sum_id^+_if_i+f^+_id_i$.  
Our DMRG results obtained for the symmetric case $E_f=0$ are summarized 
in figure~\ref{fig1} a and in figure~\ref{fig1} b where the $P_{df}=\langle d_i^{+}f_i \rangle $ expectation value 
is shown as a function of hybridization for several values of Coulomb interaction $U$ 
(figure~\ref{fig1} a) and the ratio $\Delta=P_{df}(U)/P_{df}(U=0)$ for several values of $V$
(figure~\ref{fig1} b). Figure~\ref{fig1} a clearly demonstrates that there is no nonvanishing 
$\langle d^{+}f\rangle$-expectation value in the limit of vanishing hybridization
for all examined values of $U$. At the same time, these 
data reveal an important feature of the model and namely that the 
$P_{df}$ expectation value is dramatically enhanced with increasing $U$
in comparison to the noninteracting case. 
This is  explicitly shown in figure~\ref{fig1} b where the ratio of the interacting
$P_{df}(U)$ and non-interacting $P_{df}(U=0)$ excitonic average
is plotted for several selected values of hybridization.
For all examined
values of V, the ratio $\Delta=P_{df}(U)/P_{df}(U=0)$ rapidly increases 
 with increasing interband Coulomb interaction $U$ from its 
initial value $\Delta=1$ to its saturated value $\Delta=\Delta_s$ that 
also dramatically increases with a decreasing $V$. Indeed, while 
$\Delta_s \sim 7$ for $V=0.05$ its value increases up to $\sim 200$ 
for $V=0.002$. This result is very important from the point of view 
of real rare-earth materials with $d$ and $f$ electrons. In these
materials the local hybridization is usually forbidden due to the crystal
symmetry an thus the $d-f$ coherence cannot be established. However,
according to our results, any infinitesimal hybridization, induced
by some additional mechanism, could lead to a robust excitonic 
average due to the interband Coulomb interaction. Such an additional
mechanism could be, for example, the electron-phonon interaction
$H_{\text{el-ph}}$ that can be reduced to the phonon-mediated local hybridization
(electron-electron interactions) by the standard canonical 
transformation of the form $\re^SH\re^{-S}$, where the operator $S$
is determined so that $H_{\text{el-ph}}=-[S,H_{\text{loc}}]$ and $H_{\text{loc}}$ are all
local terms corresponding to $f,d$ electrons and phonons~\cite{Menezes}.

\begin{figure}[!t]
\begin{center}
\includegraphics[width=5.8cm]{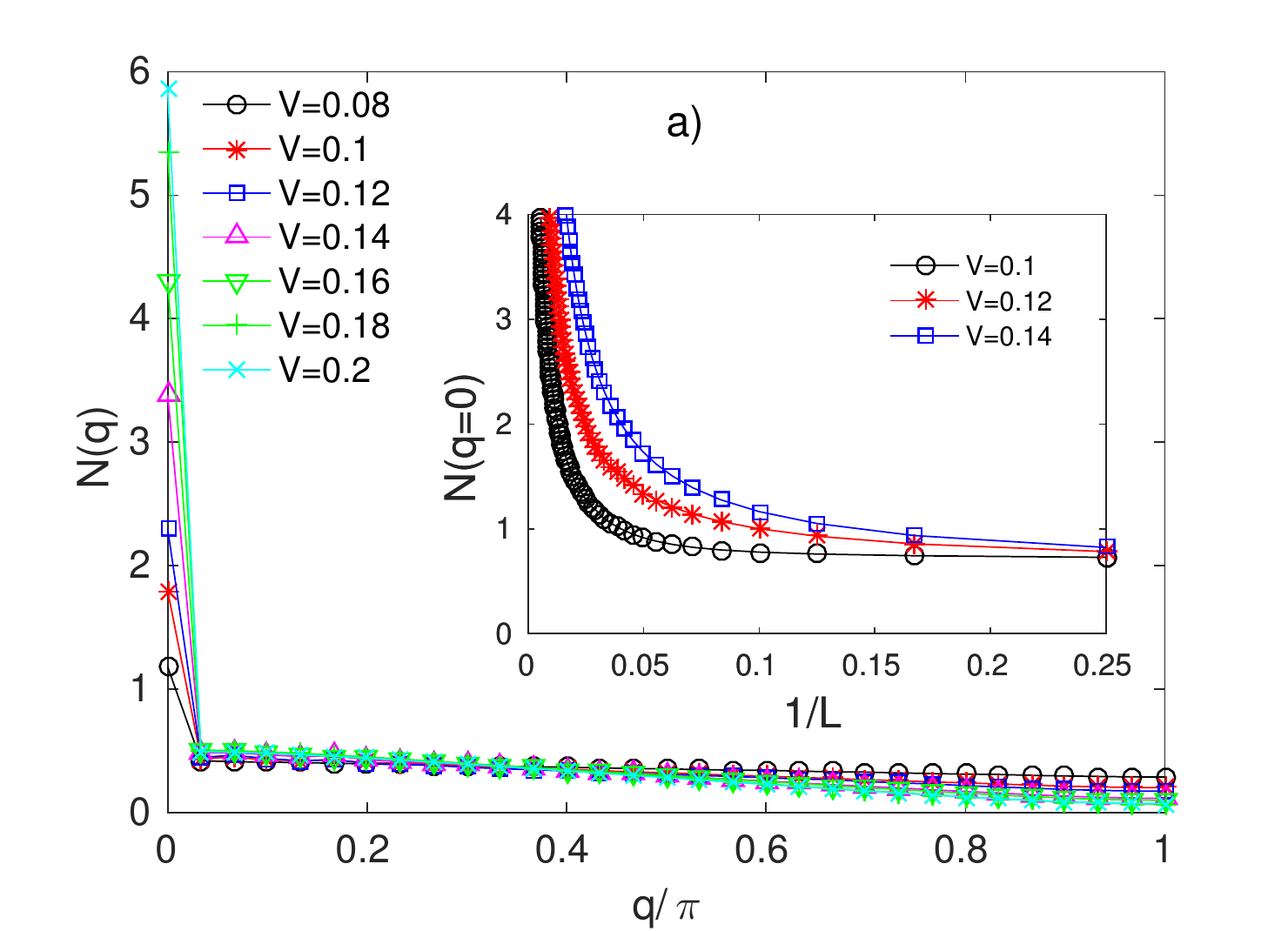}
\includegraphics[width=5.8cm]{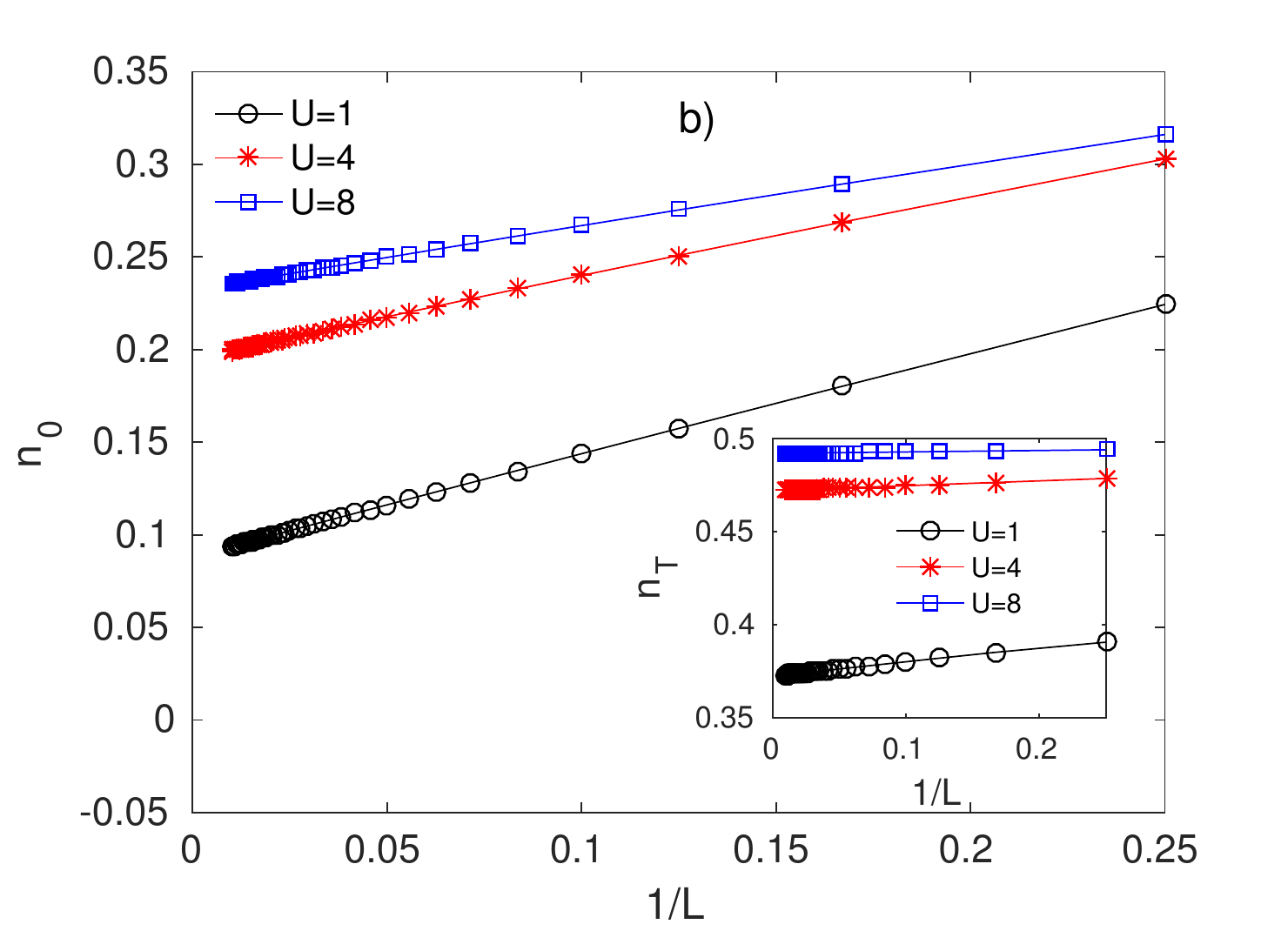}
\end{center}
%\vspace*{-0.6cm}
\caption{(Colour online) a) The excitonic momentum distribution $N(q)$
calculated for different values of $V$ at $U=1,E_f=0$ and $L=60$.
The inset shows a divergence of $N(q=0)$ for $L \to \infty$ for three 
selected values of $V$.
b) The density of zero momentum excitons $n_0$ and the total exciton density $n_T$
as functions of $1/L$ calculated for several different values of $U$ 
at $V=0.2$ and $E_f=0$~\cite{Farky_epl}.}
\label{fig2}
\end{figure}

To examine the nature of the EI state  more in detail,
we have calculated, in accordance with \cite{Kaneko2} and 
\cite{Ejima}, the exciton-exciton 
correlation function $\langle b^+_ib_j \rangle$ with $b^+_i=d^+_if_i$ and 
the excitonic momentum distribution $N(q)=\langle b^+_qb_q\rangle$ with 
$b^+_q=(1/\sqrt{L})\sum_k d^+_{k+q}f_k$.  
We have found that the exciton-exciton correlation function 
$\langle b^+_ib_j \rangle$ exhibits power-low correlations 
$|i-j|^{-\alpha}$ (with $\alpha$ between 3 and 4) and the excitonic momentum
distribution $N(q)$ diverges for $q=0$ (see figure~\ref{fig2} a), signalizing a
Bose-Einstein condensation of preformed excitons. Moreover,  
figure~2b shows that the density of zero momentum excitons 
$n_0=\frac{1}{L}N(q=0)$ as well as the total exciton density $n_T=\frac{1}{L}\sum_q N(q)$ 
 strongly depend on the values of the Coulomb interaction $U$ and that already 
for relatively small values of $U$ ($U \sim 4$) practically all particles are paired 
in electron-hole pairs with significant fraction of $n_0/n_T\sim0.5$ excitons
in the zero-momentum state. 
%

%%%%%%%%%%%%%%%%%%%%%%%%%%%%%%%%%% prb %%%%%%%%%%%%%%%%%%%%%%%%%%%%%%%%%%%%%

\subsection{Effects of $f$-electron hopping}
\begin{figure}[h!]
\begin{center}
\includegraphics[width=5.5cm]{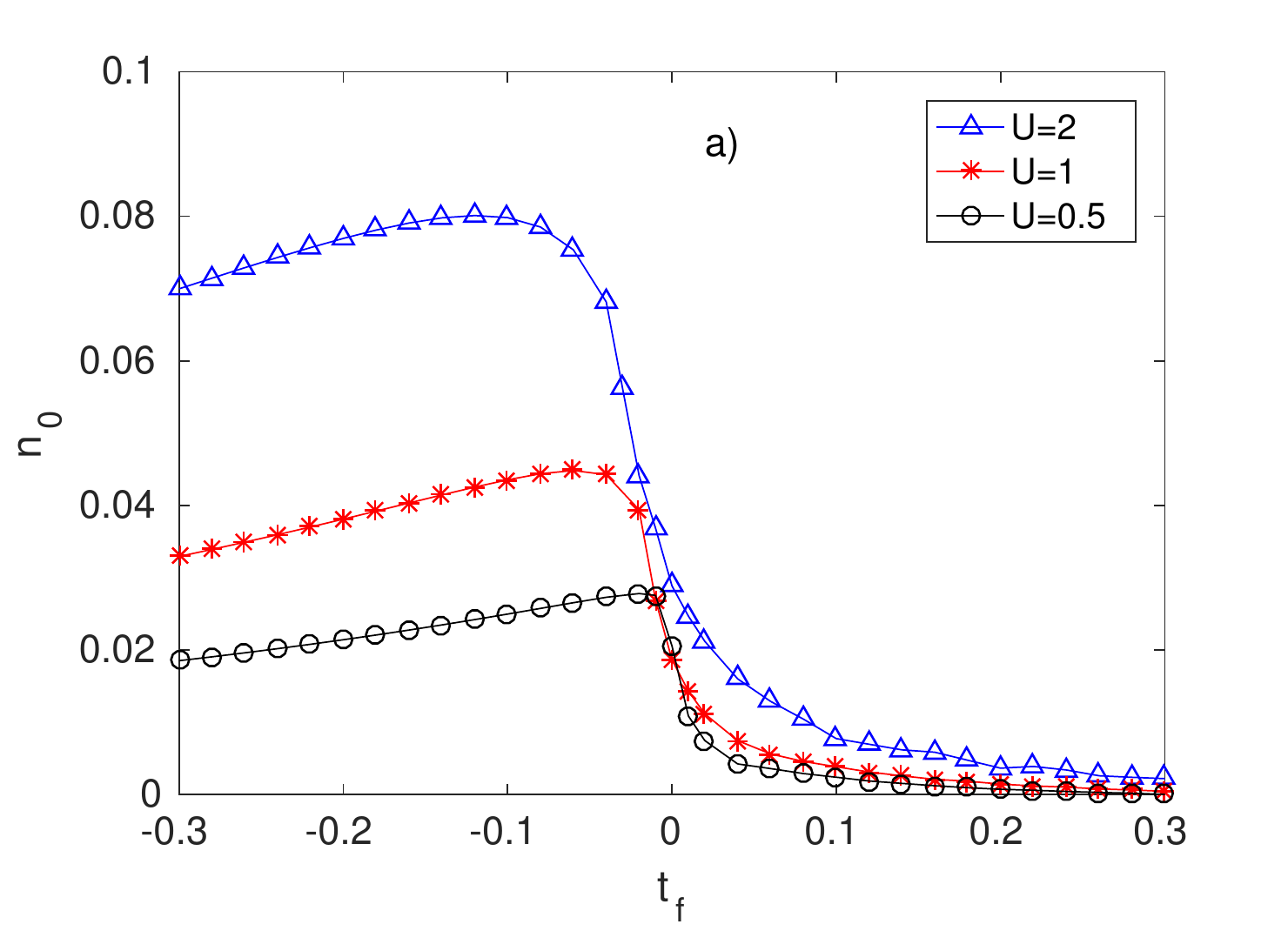}
\includegraphics[width=5.5cm]{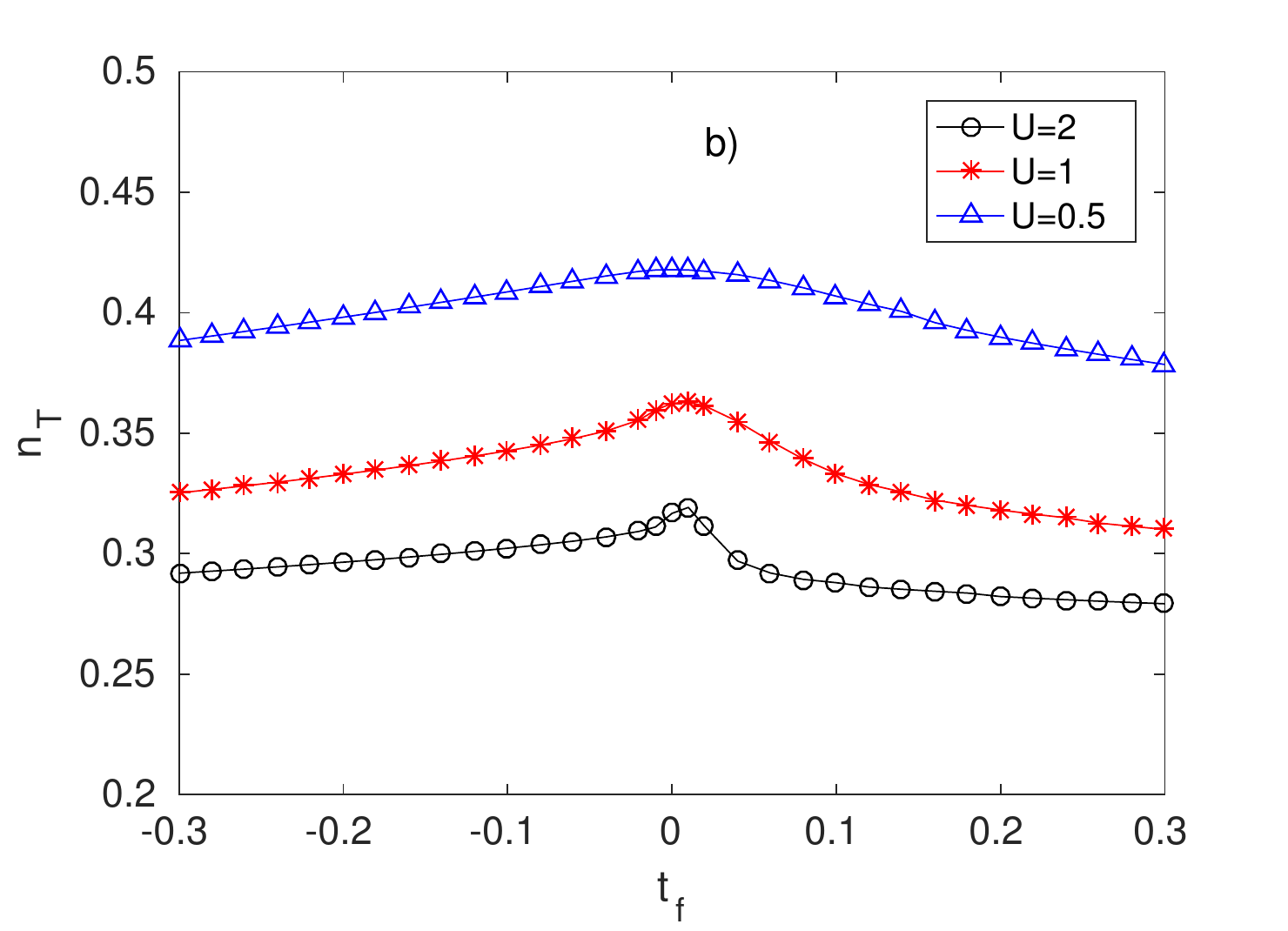}
\end{center}
%\vspace*{-0.6cm}
\caption{(Colour online) $n_0$ (a) and $n_T$ (b) as functions of $t_f$ calculated for three different values of $U$
($E_f=0$, $V=0.1$, $L=\infty$)~\cite{Farky_prb}.}
\label{fig3}
\end{figure}

With regard to the situation in real materials, where there always exists
a finite overlap of $f$ orbitals on the neighbouring sites, it is 
interesting to ask what happens if 
the $f$-electron hopping $H_{t_f}=-t_f\sum_{<i,j>}f^+_if_j$ 
is also taken into account~\cite{Farky_prb}.
In accordance with some previous theoretical studies, which 
documented strong effects of the parity of $f$ band on the stability 
of the excitonic phase~\cite{B1,B2}, we have examined the model 
for both the positive (the even parity) and negative 
(the odd parity) values of the $f$-electron hopping integrals $t_f$.
The results of our non-zero $t_f$ DMRG calculations for $n_0$ are displayed 
in figure~\ref{fig3} and they  clearly demonstrate that the zero-momentum 
condensate is suppressed in the limit of positive values of $t_f$,
while it remains robust for negative values of $t_f$.
This result is intuitively expected since our previous 
Hartree-Fock (HF) results~\cite{F3} showed that only 
the negative values of $t_f$ stabilize the ferroelectric phase, 
while the positive values stabilize the antiferroelectric phase.
The effect of $t_f$ is especially strong for $U$ small (see figure~\ref{fig3} a), 
where continuous but very steep changes of $n_0$ are observed 
for $t_f \to 0^+$.
On the contrary, the total exciton density $n_T$ (figure~\ref{fig3} b)
exhibits only a weak dependence on the $f$-electron hopping 
parameter $t_f$, over the whole interval of $t_f$ values.

\subsection{Effects of $f$-level position (pressure)}
\begin{figure}[h!]
\begin{center}
\includegraphics[width=6.0cm]{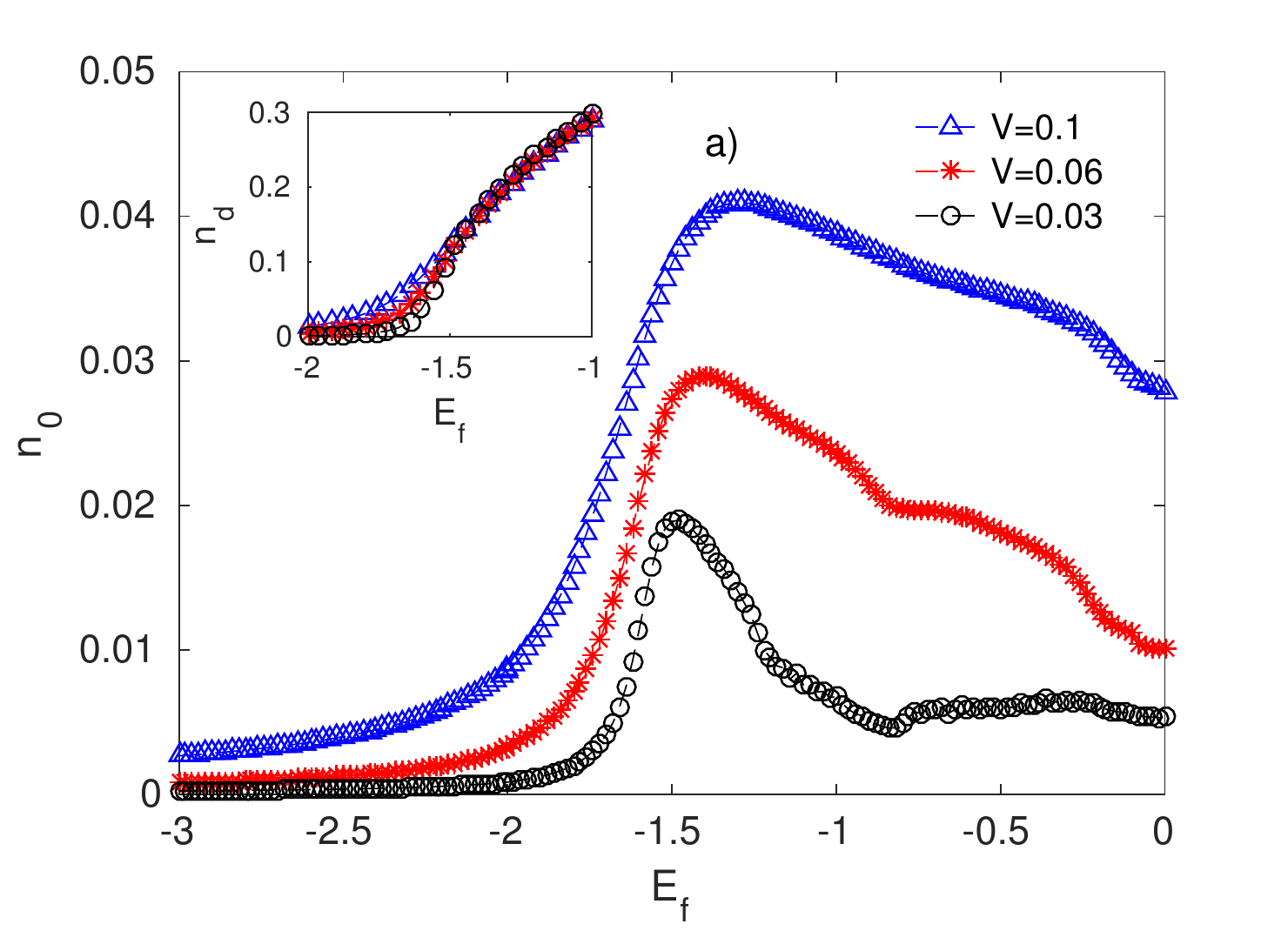}
\includegraphics[width=6.0cm]{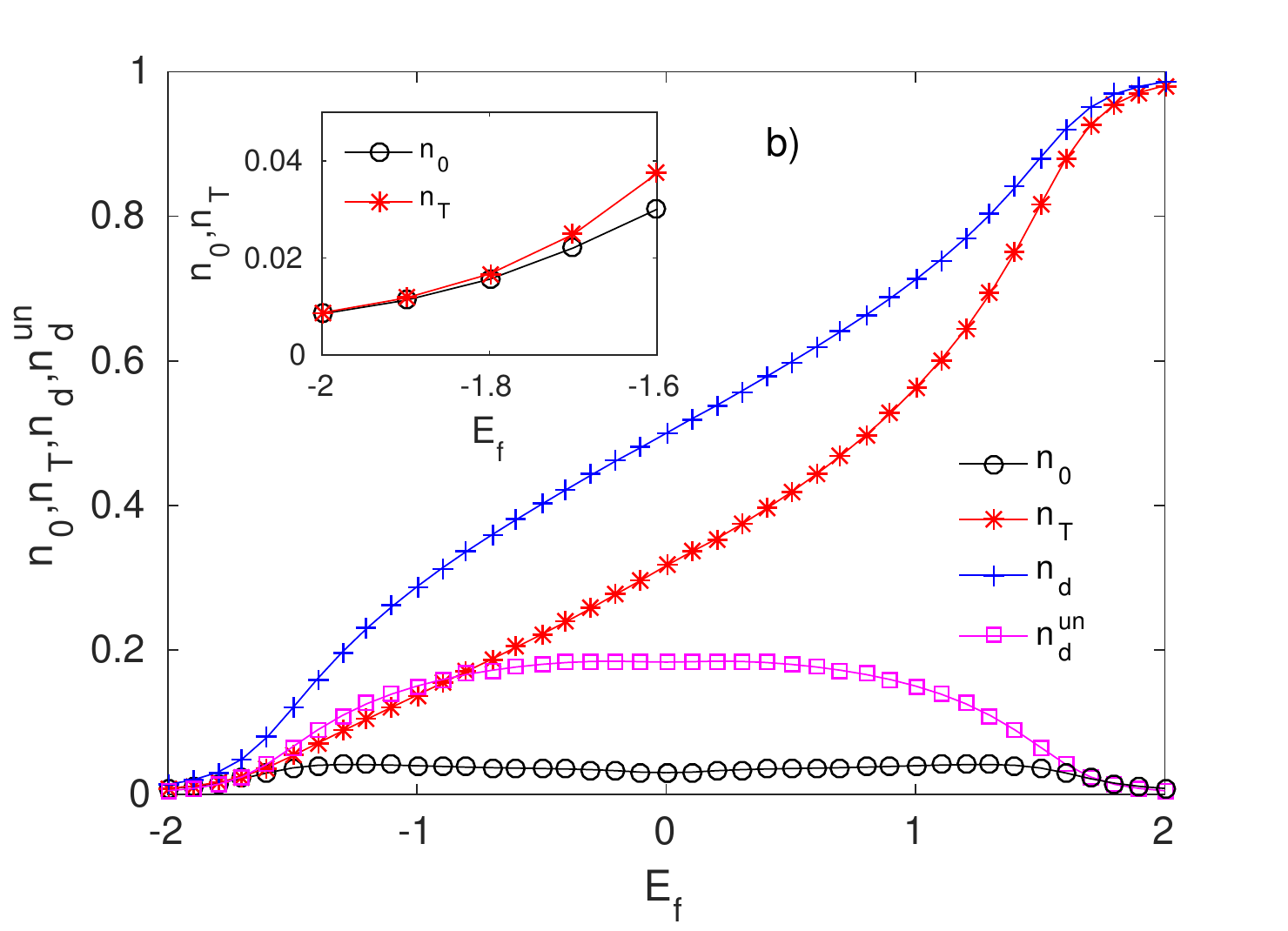}
\end{center}
%\vspace*{-0.6cm}
\caption{(Colour online) 
a) $n_0$ as a function of $E_f$ calculated  for three different values of
$V$ ($t_f=0, U=0.5, L=\infty$). The inset shows the density of $d$ electrons 
$n_d$  near $E_f=-1.5$.
b) $n_0, n_T, n_d$ and $n^{\text{un}}_d=n_d-n_T$ as functions of  $E_f$
calculated for $t_f=0, U=0.5, V=0.1$ and $L=\infty$.
The inset shows the behaviour of $n_0$ and $n_T$ near
$E_f=-2$~\cite{Farky_prb}.}
\label{fig4}
\end{figure}

So far  we have presented the results exclusively for $E_f=0$.
Let us now  briefly discuss  the effect of the change of the $f$-level
position~\cite{Farky_prb}. This study is also interesting  from the point of 
view that taking into account the parametrization between the
external pressure and the position of the $f$ level ($E_f \sim p $),
one can also deduce, at least qualitatively, their $p$ dependences from the $E_f$~dependences~of~the~ground state 
characteristics \cite{Gon}.
The resultant $E_f$ dependences of the density of zero momentum 
excitons $n_0$ are shown in figure~\ref{fig4} a
for several values of $V$ and $U=0.5$. 
One can see that the density
of zero momentum excitons is nonzero over  the whole interval
of $E_f$ values. Moreover, we have found that the values of $n_0$ 
are extremely enhanced in the region near $E_f\sim -1.5$,
which is obviously due to a significant enhancement of the 
$d$ electron population in the $d$ band (see the inset in figure~\ref{fig4} a).

To describe  the process of formation
of excitonic bound states with increasing $E_f$ more in  detail, we have also
plotted in figure~\ref{fig4} b, besides the density of zero momentum 
excitons $n_0$,  the total exciton density $n_T$,
the total $d$-electron density $n_d$
and the total density of unbond $d$ electrons $n^{\text{un}}_d=n_d-n_T$.
It is seen (see the inset in figure~\ref{fig4} b) that below $E_f \sim -1.8$,
$n_0$ and $n_T$ coincides, which means that the excitonic insulator 
in this region is practically completely driven by the condensation
of zero-momentum excitons. Above this value $n_T$ starts to 
sharply increase, while $n_0$ tends to its maximum at $E_f \sim -1.3$
and then gradually decreases to its minimum at $E_f=0$.
Similar behaviour with increasing $E_f$ also exhibits  
the density of unbond $d$ electrons $n^{\text{un}}_d$,
though the values of $n^{\text{un}}_d$ are several times 
larger than $n_0$. It is interesting to note that although 
the total exciton density $n_T$ increases over the whole 
interval of $E_f$ values, the number of unbond
$d$ electrons remains practically unchanged over the wide 
range of $E_f$ values (from $E_f=-1$ to $E_f=1$), since 
its decrease, due to the formation of excitonic pairs,
is compensated by the increase of $n_d(E_f)$.  
Thus, we can conclude that in the pressure induced case,
when the $f$-level energy shifts up with the applied 
pressure~\cite{Gon}, the model is capable of describing,
at least qualitatively, the increase in the total density of 
excitons with external pressure and the increase or decrease
(according to the initial position of $E_f$ at ambient pressure)
in the $n_0$ and $n^{\text{un}}_d$.
  
\subsection{Effects of non-local hybridization with inversion symmetry}
\begin{figure}[h!]
\begin{center}
\includegraphics[width=6.0cm]{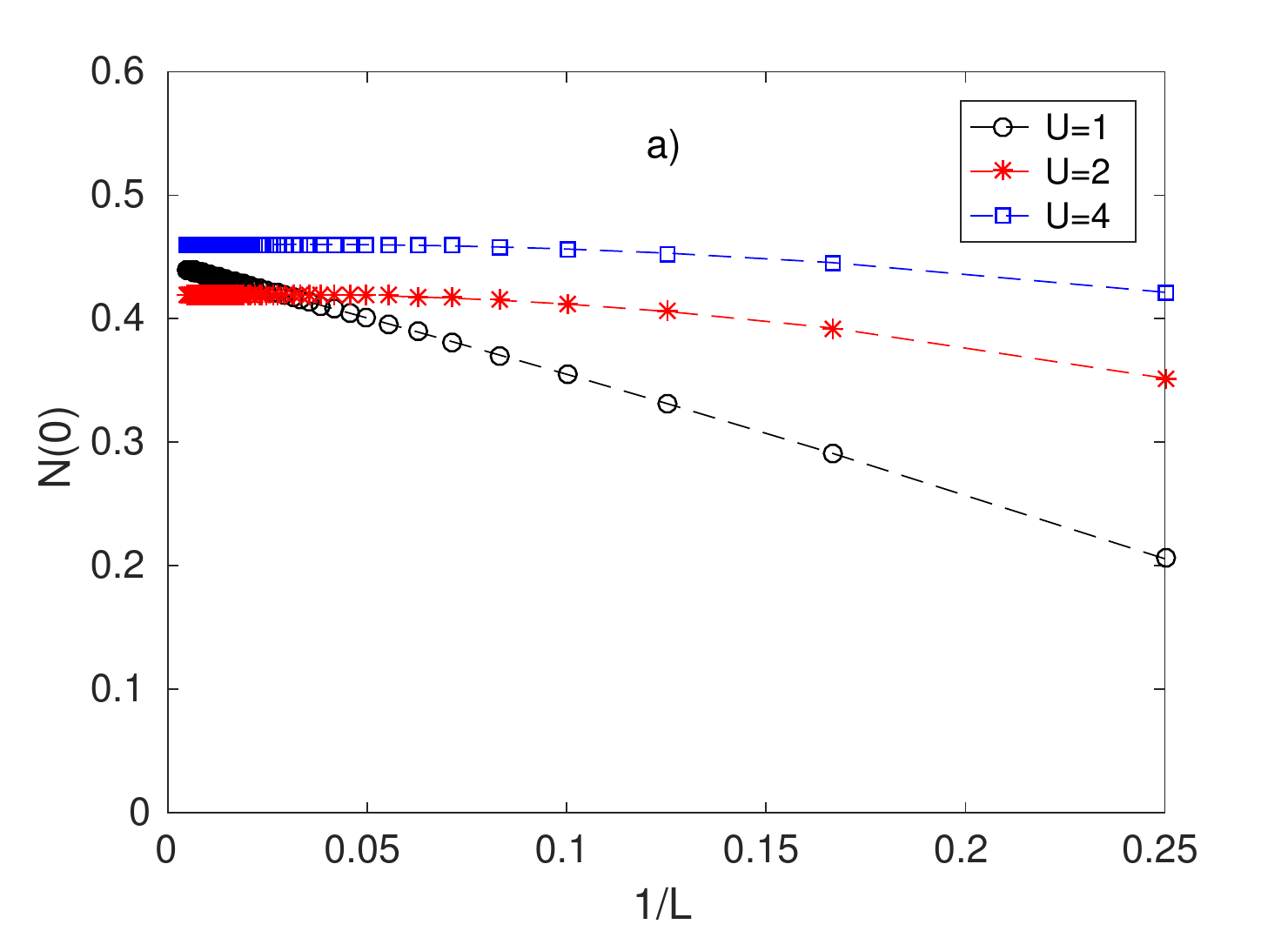}
\includegraphics[width=6.0cm]{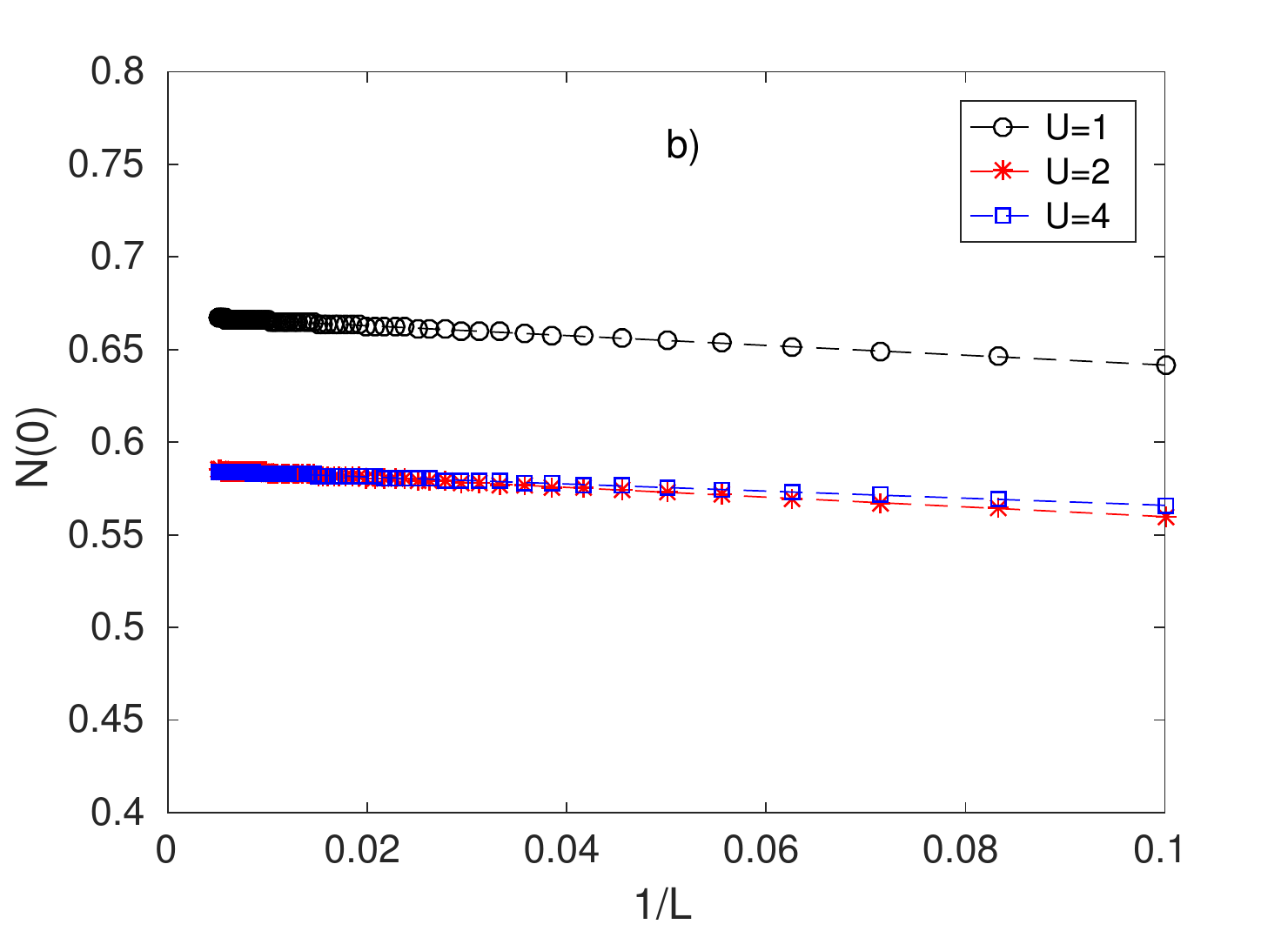}
\end{center}
%\vspace*{-0.8cm}
\caption{(Colour online)
$N(0)$ as a function of $1/L$ calculated for three different values of
$U$ and two different values of $t_f$: a) $t_f=0$, b) $t_f=-0.5$ ($E_f=0,
V_{\text{non}}=0.1$)~\cite{Farky_prb}.}
\label{fig5}
\end{figure}

As already mentioned, from the physics viewpoint, the most interesting 
case corresponds  to the case of finite non-local
hybridization~\cite{Farky_prb}. 
The importance of this term emphasizes the fact that the on-site hybridization $V$
is usually forbidden in real $d$-$f$ systems for parity
reasons. Instead of the on-site hybridization, one should consider
in these materials the non-local hybridization with 
inversion symmetry $V_{i,j}=V_{\text{non}}(\delta_{j,i-1}-\delta_{j,i+1})$
which leads to $k$-dependent hybridization of the opposite parity
that corresponds to the $d$ band [$V_k\sim \sin(k)$]~\cite{Czycholl}.
Typical examples of $1/L$ dependence of the excitonic momentum 
distribution $N(q=0)$ obtained for three representative values of 
the interband Coulomb interaction and two values of $f$-electron
hopping are displayed in figure~\ref{fig5} a and figure~\ref{fig5} b.
These results clearly demonstrate   
that there is no sign of divergence in the $1/L$-dependence of 
$N(0)$  neither for $t_f=0$ nor for $t_f=-0.05$ and thus, there 
is no signal of forming the Bose-Einstein condensate in the presence 
of non-local hybridization with the inversion symmetry.
Thus, our results indicate that the class of possible 
candidates for the appearance of the Bose-Einstein condensation 
of excitons in real $d$-$f$ materials is strongly limited,
since the local hybridization is usually forbidden in these 
systems for parity reasons and the non-local hybridization with 
the inversion symmetry does not support the formation of the 
Bose-Einstein condensate.

%%%%%%%%%%%%%%%%%%%%%%%%%%% vvn %%%%%%%%%%%%%%%%%%%%%%%%%%%%%%%%%%%%%%%%%%%%

\subsection{Combined effects of local and non-local hybridization with equal 
parity of $d$ and $f$ orbitals}
In this situation, the most
promising candidates for studying this phenomenon seem to be the systems 
with equal parity of $d$ and $f$ orbitals, where the nonlocal hybridization $H_\text{n}$ can 
be written as~\cite{Farky_ssc}: 
\begin{equation}
H_\text{n}=V_\text{n}\sum_{\langle i,j \rangle}(d^+_if_j+H.c.).
\end{equation}
In such systems, the local hybridization $V$ is allowed,
and thus one can examine the combined effects of the local and nonlocal 
hybridization within the unified picture. 
In the weak ($U<1$) and strong ($V \ll U$ and $V_\text{n} \ll U$) coupling 
limits, the model Hamiltonian $H_0+H_V+H_\text{n}$  was  
recently analyzed by Zenker et al. in~\cite{Vn}, and the corresponding mean-field 
quantum phase diagrams were presented as functions of the model 
parameters $U, V, V_\text{n}$ and $E_f$ for the half-filed band case 
$n_f+n_d=1$ and $D=2$. Moreover, examining the effects of the local 
$V$ and nonlocal $V_\text{n}$ hybridization, they found that in the 
pseudospin space ($c^{+}_{i\uparrow}=d^+_i$,$c^{+}_{i\downarrow}=f^+_i$), 
the nonlocal hybridization $V_\text{n}$ favors the staggered Ising-type ordering 
along the $x$ direction, while $V$ favors a uniform polarization 
along the $x$ direction and the staggered Ising-type ordering  
along the $y$ direction. In our paper~\cite{Farky_ssc} we have examined the model for arbitrary $V$ and 
$V_\text{n}$ and unlike the paper of Zenker et al.~\cite{Vn}  we have focused 
our attention primarily on a description of process of formation 
and condensation of exitonic bound states. 

\begin{figure}[!t]
\begin{center}
\includegraphics[width=6.0cm]{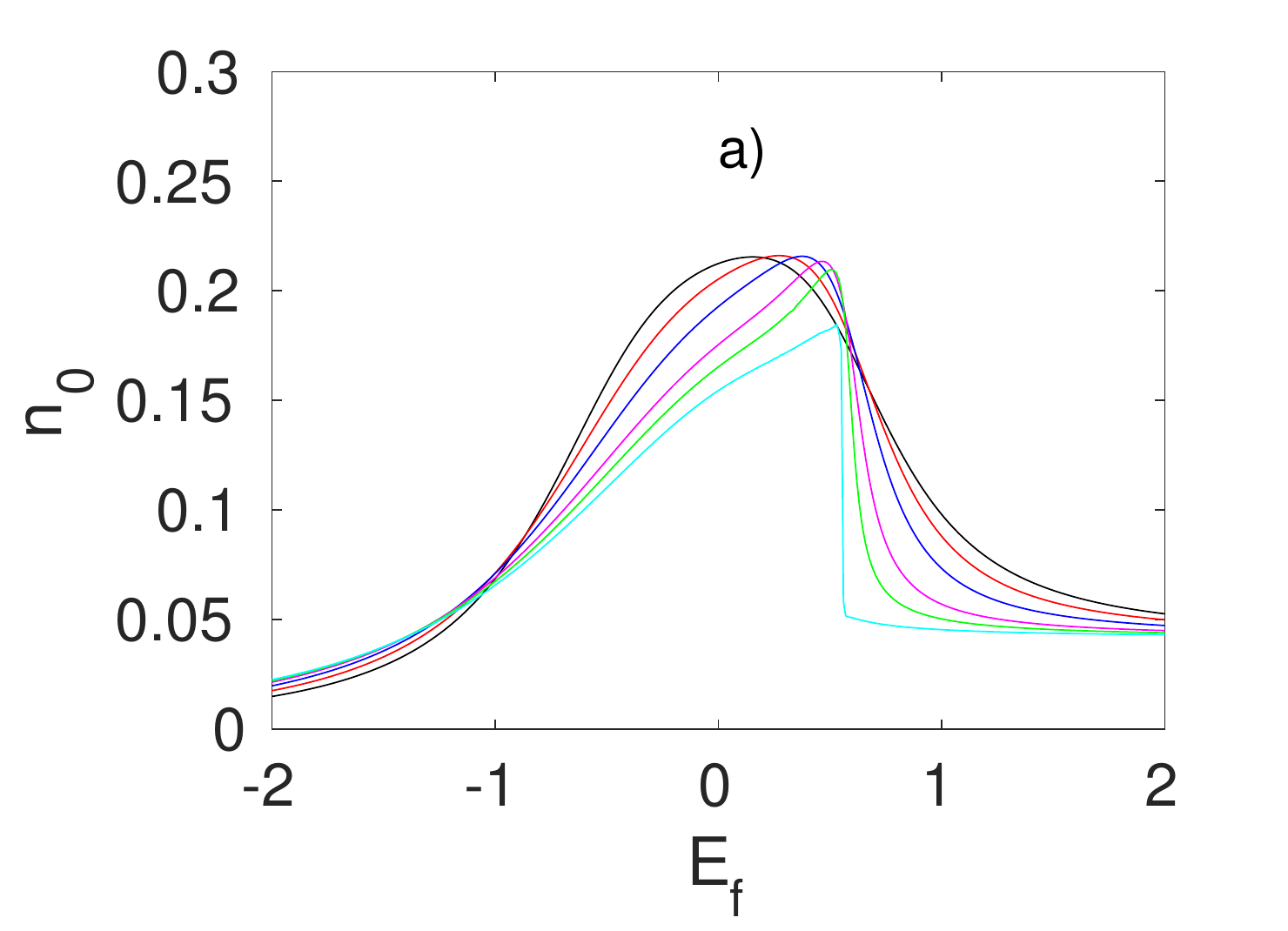}
\includegraphics[width=6.0cm]{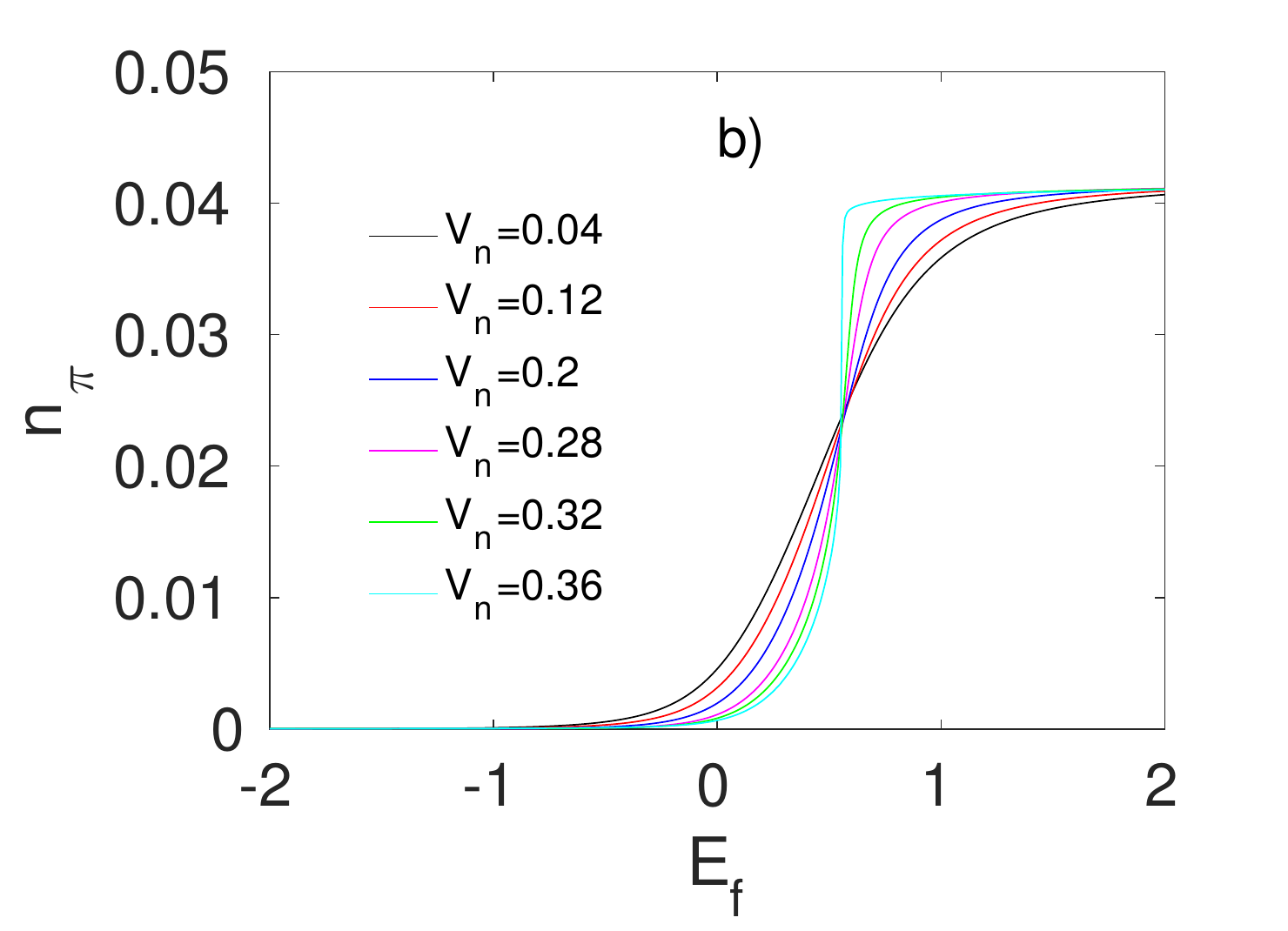}
\includegraphics[width=6.0cm]{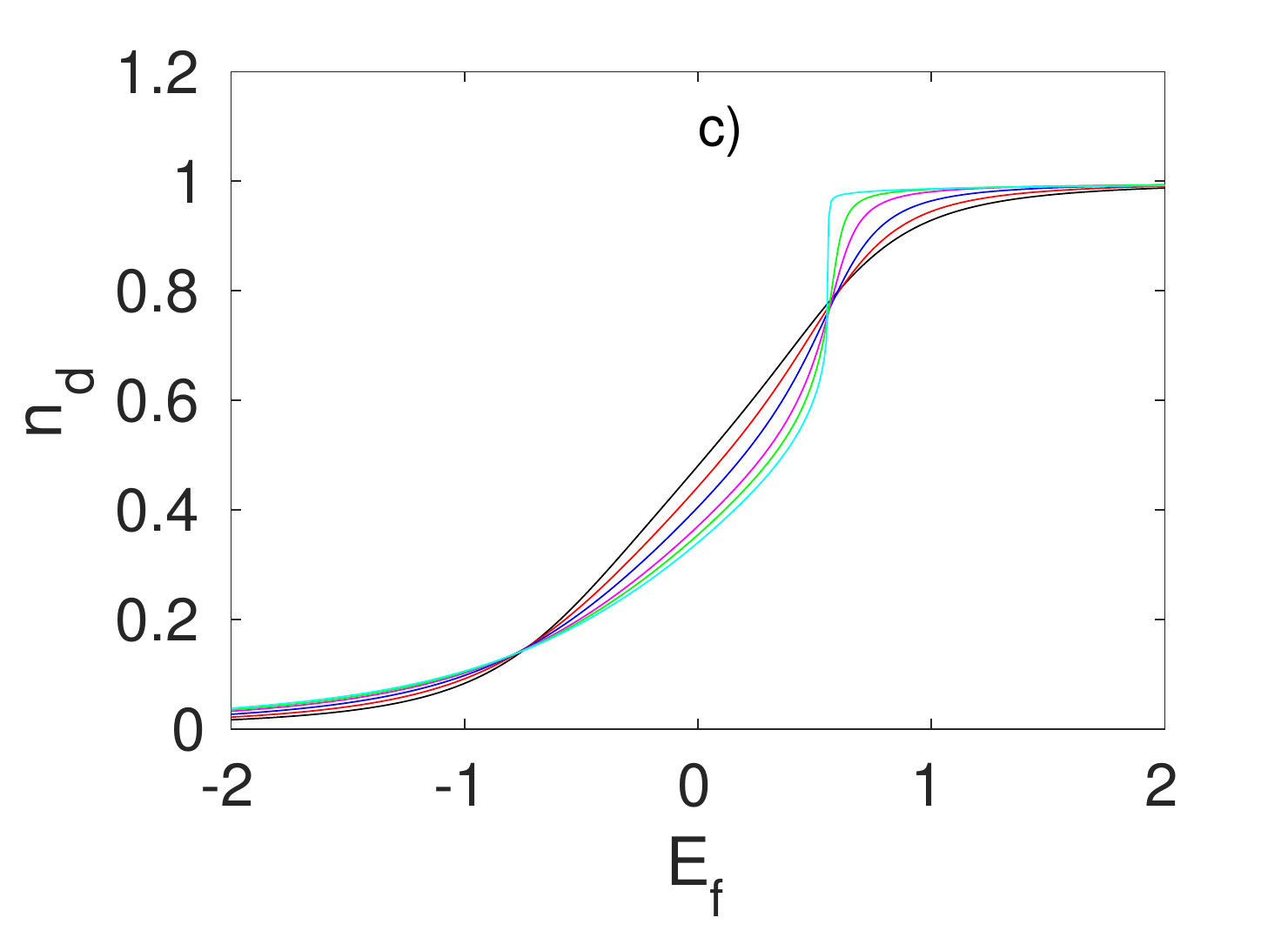}
\includegraphics[width=6.0cm]{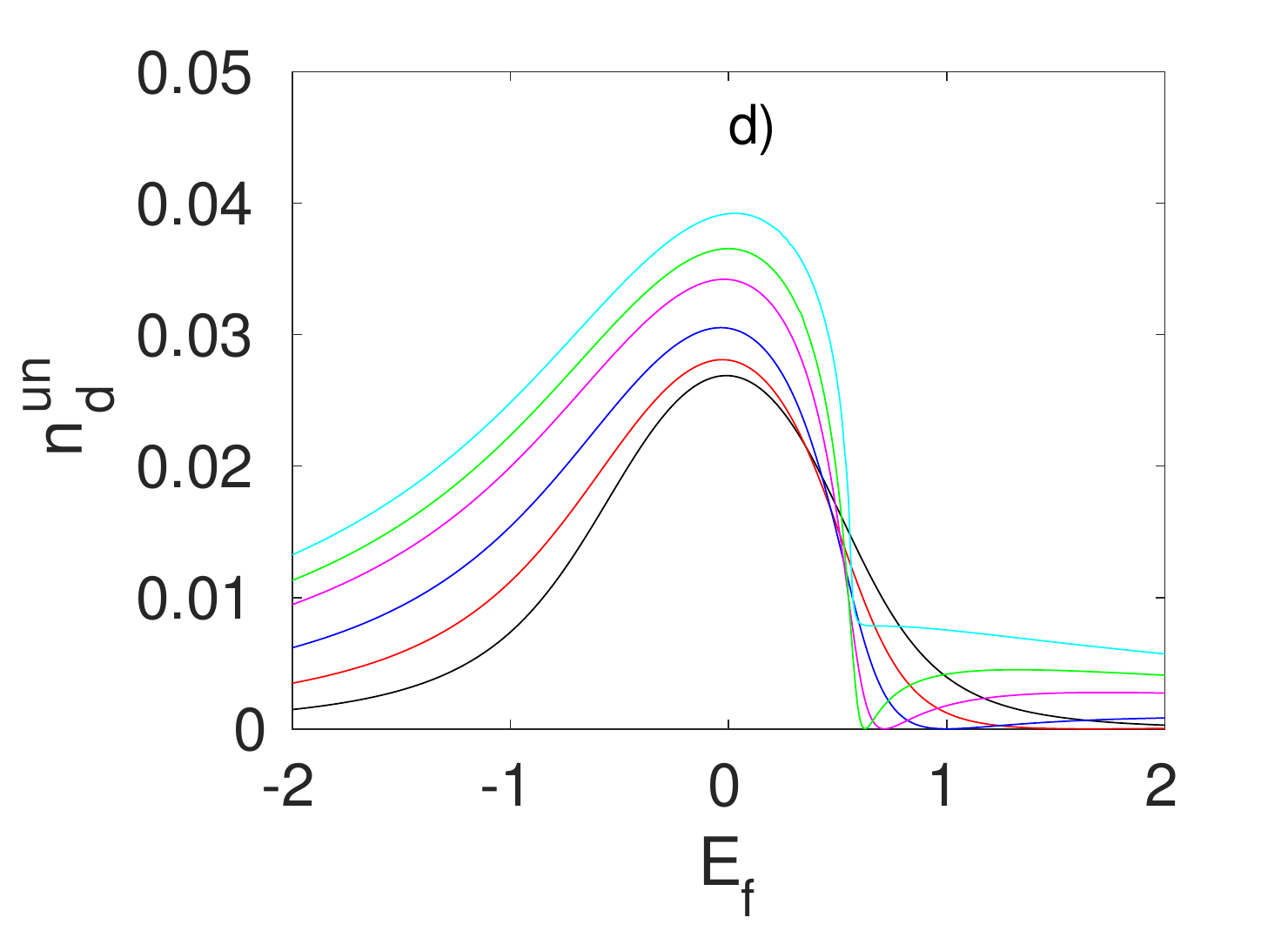}
\end{center}
%\vspace*{-0.8cm}
\caption{(Colour online) The density of zero-momentum excitons $n_0$ (a),
the density of $\piup$-momentum excitons $n_{\piup}$ (b),
the total $d$-electron density $n_d$ (c), and the total density of unbound 
$d$ electrons $n^{\text{un}}_d=n_d-n_T$ (d) as functions of $E_f$ 
calculated for $U=4, V=0.2, L=60$ and six different values of
$V_\text{n}$~\cite{Farky_ssc}.}
\label{fig6}
\end{figure}

Let us discuss the results obtained for $n_0=\frac{1}{L}N(q=0)$,
$n_{\piup}=\frac{1}{L}N(q=\piup)$, $n_d$ and $n^{\text{un}}_d$ as functions 
of the $f$-level position $E_f$  which can give us, at least qualitatively, 
the answer to the very important question, and namely, how these quantities 
change with the applied pressure $p$. 
In figure~\ref{fig6} we present the resultant behaviours of $n_0, n_{\piup}, n_d, n^{\text{un}}_d$ 
as functions of the $f$-level position $E_f$ obtained by the DMRG method for 
$V=0.2$ and several different values of $V_\text{n}$. 
In all examined cases, the density of zero-momentum excitons is 
the most significantly enhanced for $d$-electron densities near 
the half-filled band case $E_f=0$ and $n_d = 1/2$. The changes of 
$n_0$ are gradual for $E_f<0$ and very steep, but still continuous,
for $E_f >0$. The fully different behaviour 
exhibits the density of $\piup$-momentum excitons $n_{\piup}$. Its
enhancement with increasing $E_f$ is practically negligible 
for $E_f < 0$, but from this value $n_{\piup}$ it starts to  
sharply increase and tends to its saturation value corresponding to the
fully occupied $d$ band $n_d \sim 1$. The density of unbound $d$ 
electrons $n^{\text{un}}_d$  exhibits a very
simple  behaviour for $E_f <0$. In this limit, $n^{\text{un}}_d$ gradually increases 
with increasing $E_f$ for all examined values of nonlocal hybridization 
$V_\text{n}$. However, in the opposite case ($E_f > 0$), the density of unbound
$d$ electrons $n^{\text{un}}_d$  behaves fully differently for $V_\text{n} < V^c_n$
and $V_\text{n} > V^c_n$, where $V^c_n \sim 0.2$. For $V_\text{n} < V^c_n$, 
the density of unbound $d$ electrons $n^{\text{un}}_d$ gradually decreases 
with an increasing $E_f$ and tends to zero when $E_f$ approaches the 
upper edge of the noninteracting band $E_f=2$,
but in the opposite limit the  density of unbound $d$ electrons 
$n^{\text{un}}_d$ decreases by the interval of $E_f$ values from $E_f=0$~to~$E^c_f(V_\text{n})$,~and $n^{\text{un}}_d$ starts to increases again for 
$E_f > E^c_f(V_\text{n})$.
\begin{figure}[!t]
\begin{center}
\includegraphics[width=7.0cm]{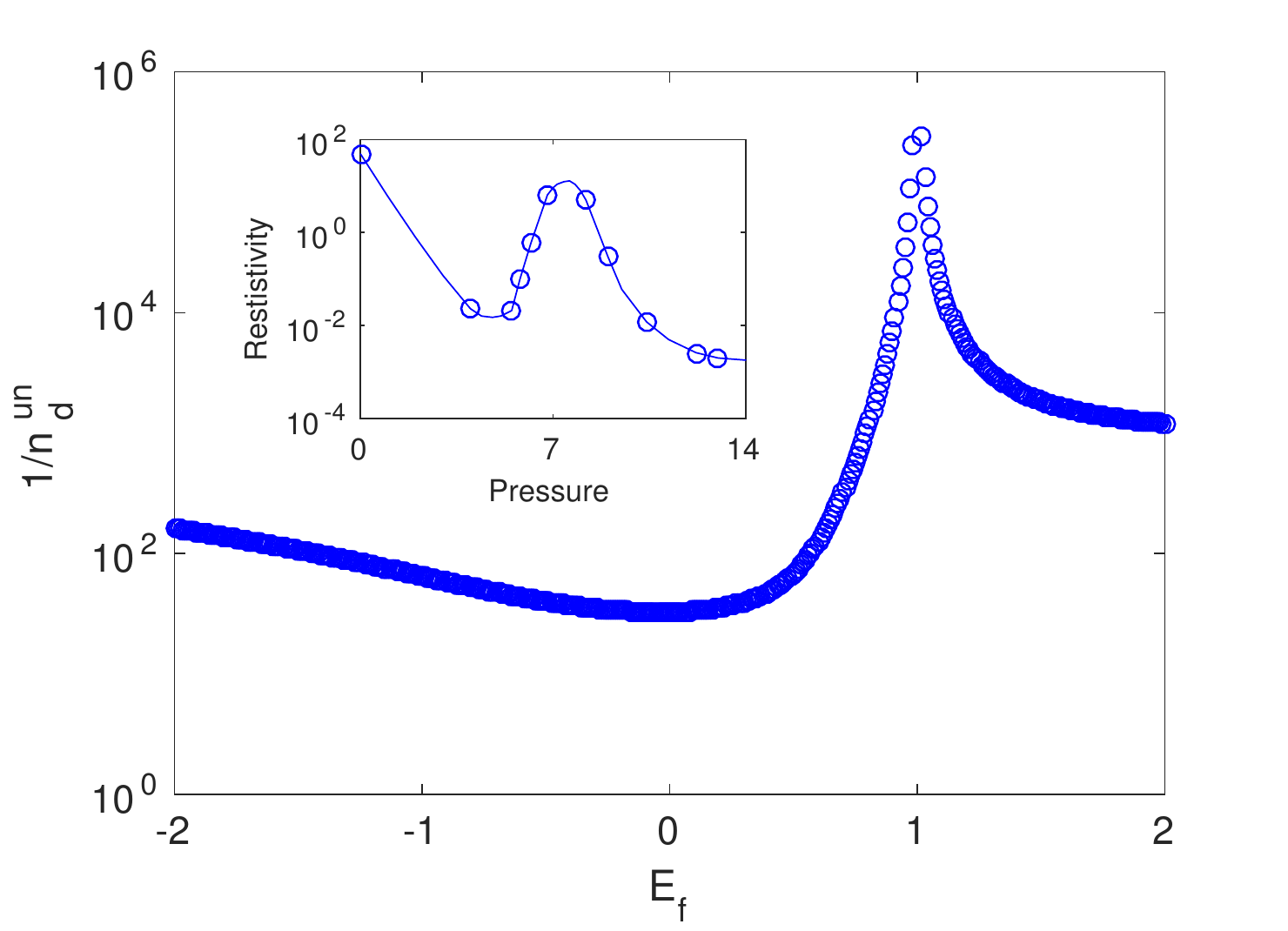}
\end{center}
%\vspace*{-0.8cm}
\caption{(Colour online) The inverse value of the density of unbound $d$-electrons 
$n^{\text{un}}_d$ as a function of the $f$-level energy $E_f$ calculated for 
$U=4, V=0.2, V_\text{n}=0.2$ and $L=\infty$~\cite{Farky_ssc}. The inset shows the resistivity 
as the function of pressure in TmSe$_{0.45}$Te$_{0.55}$ at
4.2~K~\cite{Wachter}.}
\label{fig7}
\end{figure}
Taking into account the above mentioned 
parametrization between $E_f$ and the external pressure $p$, as well as 
the fact that the electrical conductivity is proportional to the density 
of unbound electrons  $n^{\text{un}}_d$ (and the electrical resistivity to  $1/n^{\text{un}}_d$), 
the results discussed above could have very important physical 
consequences. Indeed, in figure~\ref{fig7} we have plotted the  quantity 
$1/n^{\text{un}}_d$ (in the logarithmic scale) as a function of $E_f$ and compare 
it with experimental measurements of the pressure dependence of the 
electrical resistivity in the mixed valence compound TmSe$_{0.45}$Te$_{0.55}$
(see the inset in figure~\ref{fig7}). One can see that there is a nice qualitative 
accordance between our theoretical predictions and experimental results 
of Wachter et al.~\cite{Wachter}. In spite of the fact that our model is in many
aspects very simplified, the physics that could lead to the unusual behaviour  of the \newline electrical resistivity in TmSe$_{0.45}$Te$_{0.55}$ under the external 
pressure seems to be clear. This is a result of the  formation and condensation 
of excitonic bound states of conduction-band electrons and valence-band holes.

%%%%%%%%%%%%%%%%%%%%%%%%%%% fkm_ch23%%%%%%%%%%%%%%%%%%%%%%%%%%%
\subsection{Effects of non-local Coulomb interactions}
The above discussed results show that the Falicov-Kimball model has a great potential 
to describe some of the anomalous features of real complex materials
such as rare-earth compounds. On the other hand, it should be noted
that the original version of the model, as well as its extensions
discussed above, represent a too crude approximation of real rare-earth
compounds, since we neglect all nonlocal Coulomb interactions, that
can change this picture. For a correct description of these materials
one should take into account at least the following nonlocal
Coulomb interaction terms~\cite{Farky_epjb}:
\begin{equation}
H_{\text{non}}=
U_{dd}\sum_{<ij>}n^d_in^d_j+
U_{df}\sum_{<ij>}n^d_in^f_j+
U_{ff}\sum_{<ij>}n^f_in^f_j+
U_{ch}\sum_{<ij>}d^+_id_j(n^f_i+n^f_j),
\label{eq3}
\end{equation}
which represent  the nearest-neighbour Coulomb interaction 
between two $d$ electrons (the first term), between one $d$ and one $f$ 
electron (the second term), between two $f$ electrons (the third term)
and the so-called correlated hopping (the last term).

\begin{figure}[!t]
\begin{center}
\includegraphics[width=6.0cm]{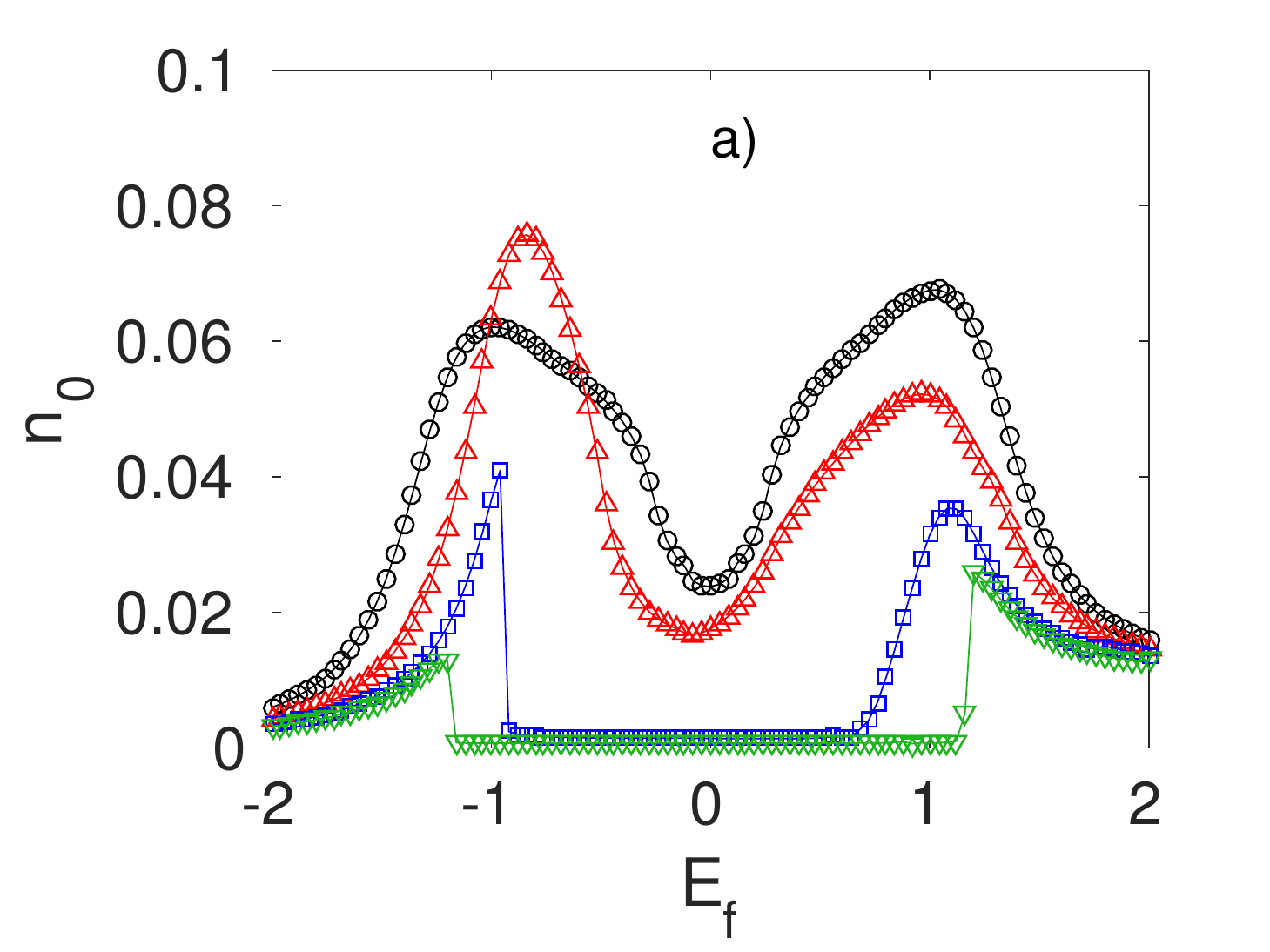}
\includegraphics[width=6.0cm]{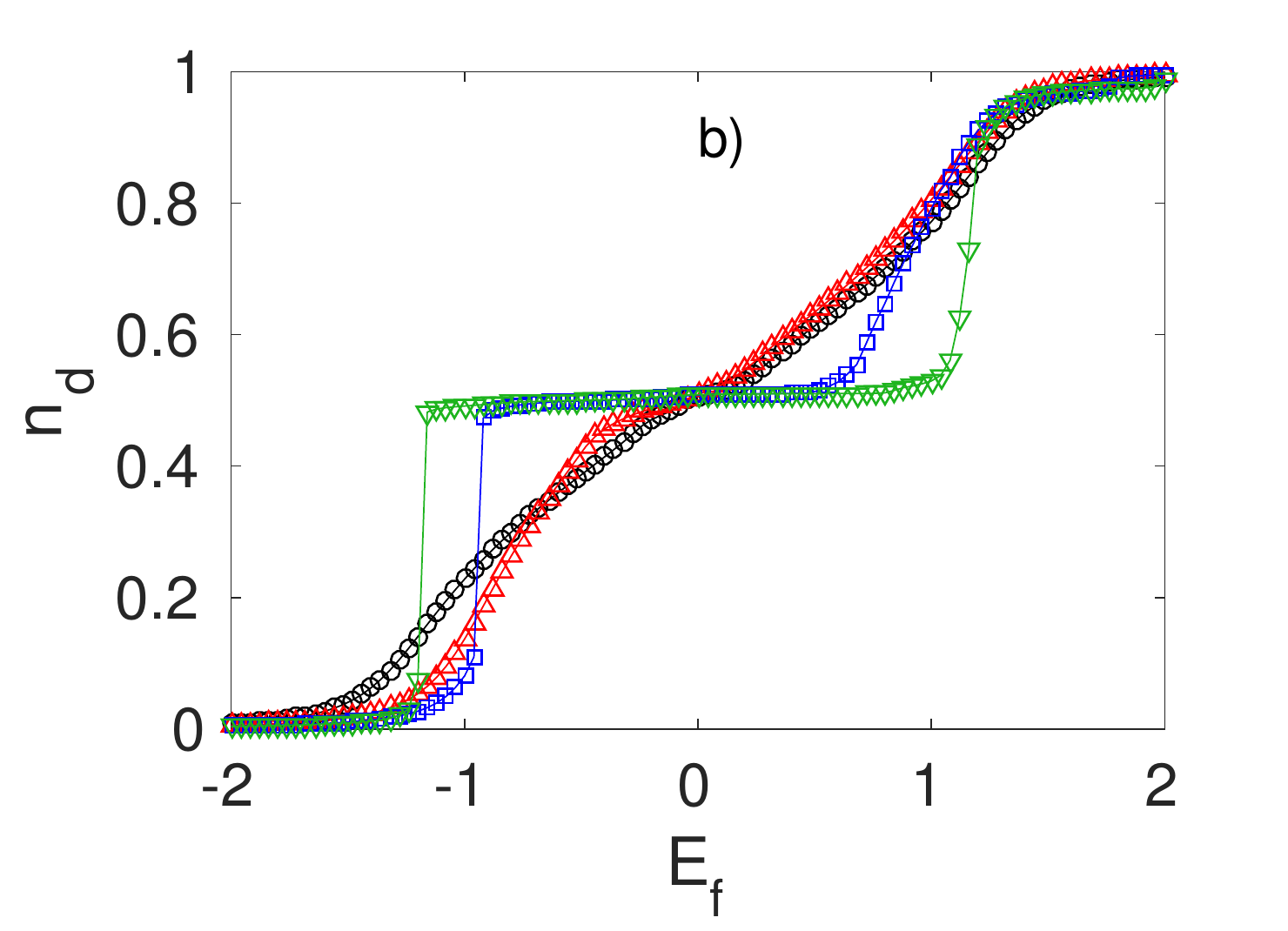}
\includegraphics[width=6.0cm]{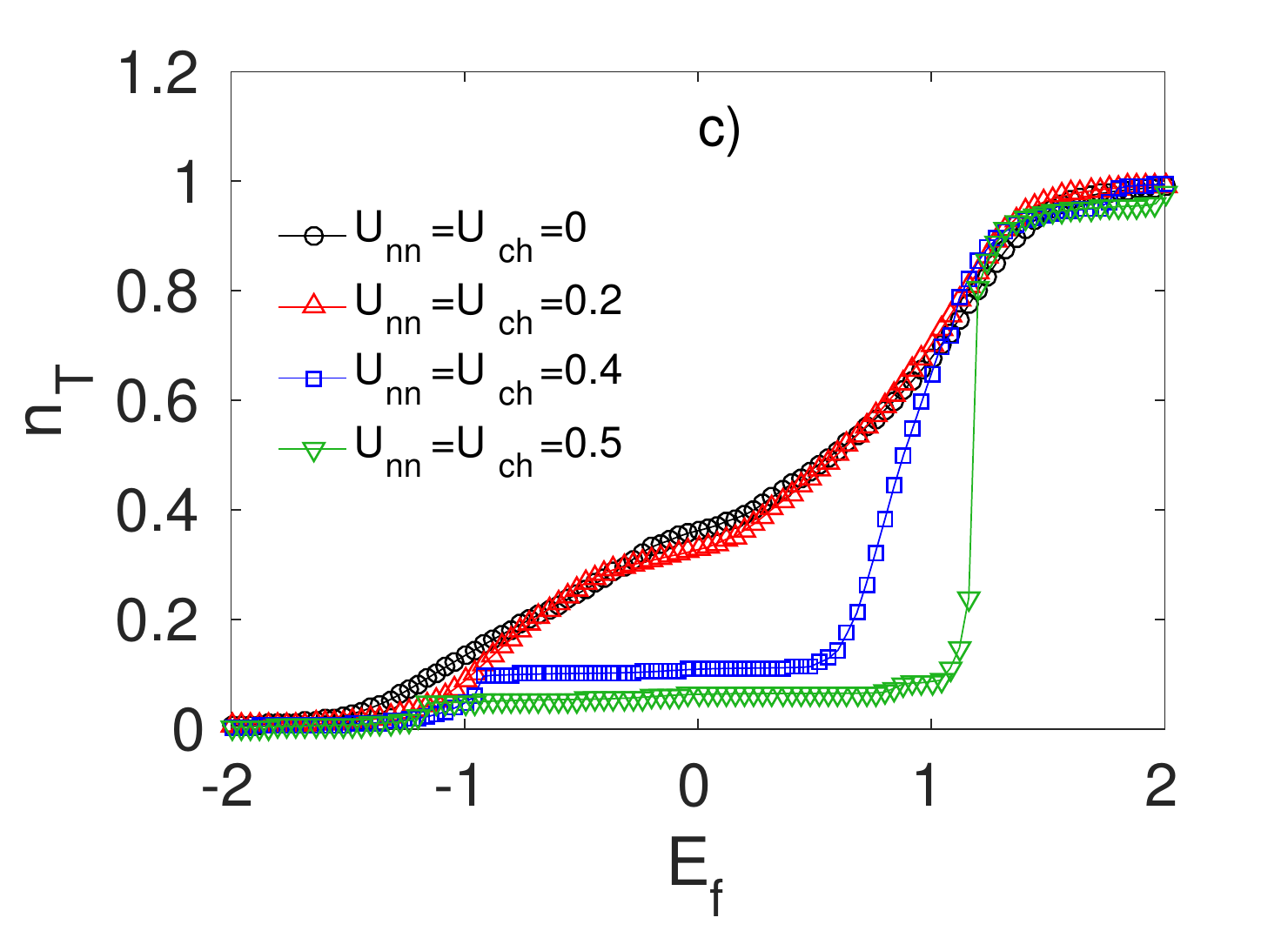}
\includegraphics[width=6.0cm]{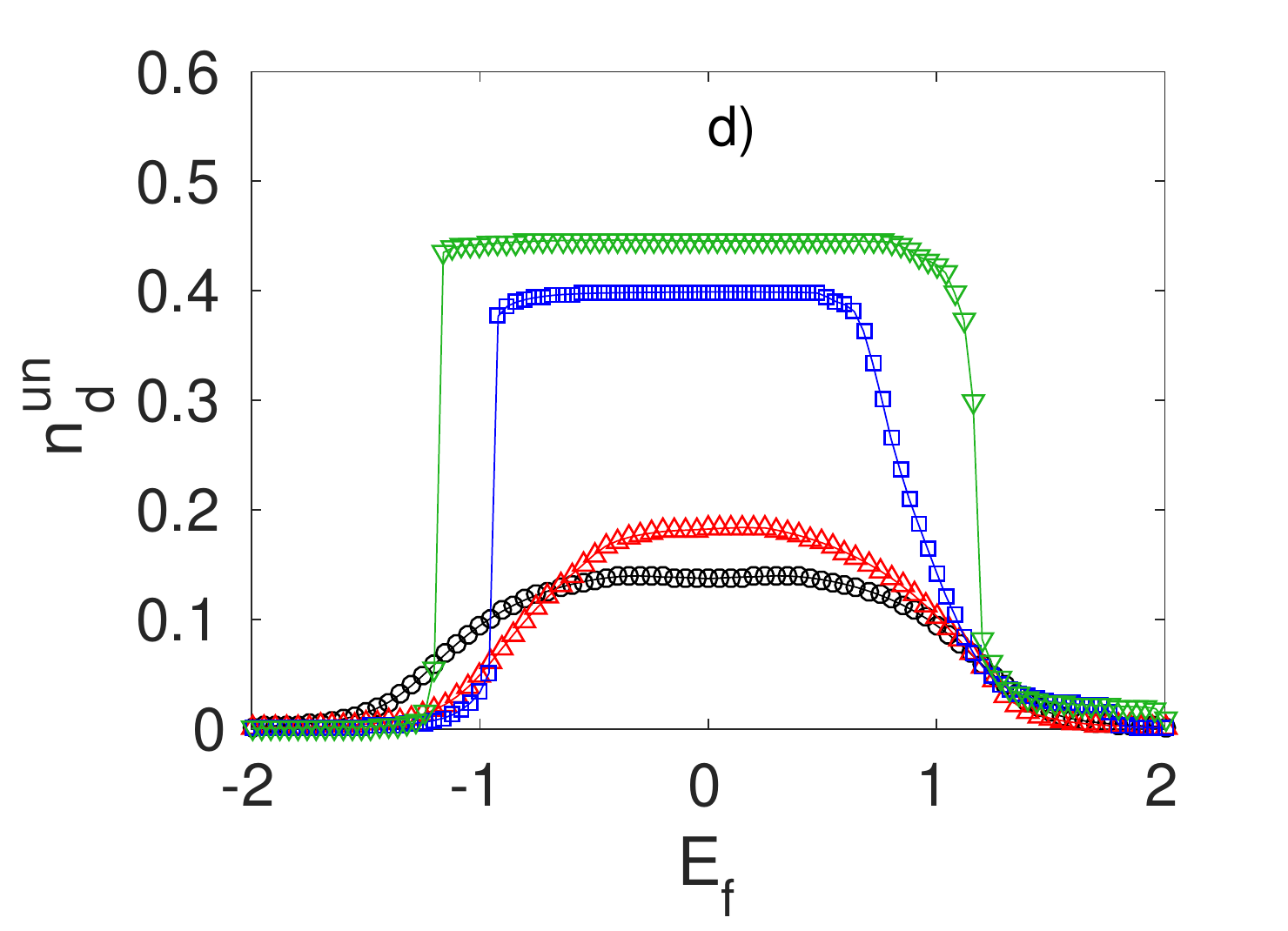}
\end{center}
%\vspace*{-0.8cm}
\caption{(Colour online) $n_0, n_d, n_T$ and $n^{\text{un}}_d=n_d-n_T$ as functions of $E_f$
calculated for four different values of $U_{ch}$ ($U_{ch}=0, 0.2, 0.4, 0.5$)
at $U_{nn}=U_{ch}, U=1, V=0.1, L=100$ and $n_f+n_d=1$~\cite{Farky_epjb}.}
\label{fig8}
\end{figure}

There is a number of papers, were the influence of individual
interaction terms from (\ref{eq3}) on the ground state properties
of the Falicov-Kimball model has been studied. 
However, there are only a few where the combined 
effects of two or three terms were considered. 
Among the papers dealing with the influence of individual
interactions, let us mention the work~\cite{Fark_aps} (and
references therein) where the effects of nonlocal interaction between
$d$ and $f$ electrons are examined and the excellent papers
of Shvaika et al.~\cite{Shvaika1,Shvaika2,Shvaika3} where rigorous results 
for the influence of the correlated hopping on  the thermodynamical functions 
were derived within the local approach and then used for a description of various 
physical problems. Among the papers dealing 
with combined effects of two or three terms, let us mention 
the works~\cite{Fark_app,Lemanski} (and references therein). 
From this point of view, the model Hamiltonian
$H=H_0+H_V+H_{\text{non}}$ considered here represents one of the most
complex extensions of the Falicov-Kimball model used for a description
of ground state properties of strongly correlated systems.   
Here, we focus our attention exclusively on a discussion of two 
main problems, and namely, the process of formation and condensation 
of excitonic bound states and the problem of valence transitions 
in the generalized Falicov-Kimball model. To simplify numerical calculations, 
we adopt here the following model $U_{dd}=U_{ff}=U_{df}=U_{nn}$, that allows us   
to reduce the number of model parameters and at the same time to keep
all nonlocal interaction terms nonzero. The physically most interesting case corresponds 
to the situation where both ($U_{nn}$ as well as $U_{ch}$)
interactions are switched on simultaneously and numerical results for this
case are summarized in figure~\ref{fig8}. 
One can see that combined effects of non-local interactions lead to a number
of interesting results: 
(i)  strong suppression of the zero-momentum condensate in the region of $E_f$,  
where $n_d \sim 0.5$, (ii)  stabilization of the intermediate phase
with $n_d \sim 0.5$ for increasing $U_{nn}=U_{ch}$, (iii)  
strong enhancement of the total density of unbond $d$ electrons
$n^{\text{un}}_d$ with an increase of $U_{nn}=U_{ch}$. (iv)
 stabilization of zero momentum condensate   
for some values of the $f$-level energy $E_f$ in the 
weak coupling limit $U_{nn}=U_{ch} \sim 0.2$, (v)  appearance 
of  discontinuous valence transitions for sufficiently large 
values of $U_{nn}=U_{ch} \sim 0.4$ and (vi)  discontinuous
disappearance of the density of zero momentum excitons,
as well as  discontinuous changes in the total density 
of excitons $n_T$ and the total density of unbond $d$ 
electrons $n^{\text{un}}_d$ at the valence transition points.

The appearance of  discontinuous changes in some ground-state
observables such as the density of conduction $d$ (valence $f$) electrons,
the density of zero-momentum condensate, the density of unbond
electrons, is a very important result from the point of view of 
rare-earth compounds. In some of them, e.g., the mixed valence 
system SmS such discontinuous changes are experimentally observed  
in the density of valence electrons when  the external hydrostatic 
pressure is applied~\cite{SmS}, though they were not 
satisfactorily described so far. Indeed, as mentioned
above, the SmS compound is a mixed valence system, with 
fluctuating valence and thus for its description one should take
into account the hybridization between the localized $f$ and 
conduction $d$ electron states. However, more reliable methods,
such as alloy-analog approximation~\cite{AAA}, renormalization group 
method~\cite{Hanke}, exact diagonalization method~\cite{Fark_zphys}, predict
only the continuous valence transitions within the Falicov-Kimball
model extended by the local hybridization. Here, we show that considering
the parametrization between the external pressure $p$ and the
$f$-level position $E_f$, the pressure induced discontinuous
valence transitions are possible to generate also in such a system
under a very realistic assumption,  namely, that nonlocal interactions
are switched on.  This opens up a new route to the understanding of various
ground-state anomalies observed in the rare-earth compounds
within the unified picture. 

Finally, it should be noted that although all the results presented in this
review were obtained for the one-dimensional case, their validity is
probably much more general. Indeed, a direct comparison of our
one-dimensional DMRG and two dimensional Hartree-Fock
results~\cite{Farky_prb}, obtained for the density of zero-momentum 
excitons as a function of $t_f$ and $E_f$, revealed only a weak dependence 
of $n_0$ on the system dimension indicating a possible extension 
of our one-dimensional DMRG results to real two and three dimensional systems.       
Moreover, in the two-dimensional case, we can switch off completely the local 
hybridization, since in this case the excitonic condensate can be generated by other 
terms (the $f$-electron hopping), modelling more realistically the situation
in rare-earth compounds.

\section*{Acknowledgements}
 This work was supported by projects VEGA 2-0112-18, APVV-17-0020, 
ITMS 2220120047, ITMS 26230120002 and IMTS 26210120002.

\ukrainianpart

\title{Дослідження конденсації екситонів методом DMRG для узагальненої моделі Фалікова-Кімбала}
%DMRG study of exciton condensation in the extended Falicov-Kimball model
%
\author{П. Фаркашовський}
\address{
 Інститут експериментальної фізики, Словацька академія наук, вул. Ватсонова 47, Кошиці, Словаччина
}

\makeukrtitle

\begin{abstract}
Формування і конденсація зв’язаних екситонних станів між електронами з зони провідності та дірками з валентної зони, безумовно, належить до однієї з найбільш захоплюючих ідей сучасної фізики твердого тіла. У цьому короткому огляді, ми представляємо останній прогрес у цій галузі, що був досягнутий завдяки розрахункам методом ренорм групи матриці густини (DMRG) для різних узагальнень моделі Фалікова-Кімбала. Особлива увага приділяється опису найважливіших механізмів (взаємодій), які впливають на стабільність екситонної фази, а саме: (i) міжзонна $d$-$f$ кулонівська взаємодія, (ii) перенос $f$-електронів, (iii) парна і непарна нелокальна гібридизація, (iv) комбіновані ефекти локальної та нелокальної гібридизації, (v) кулонівська взаємодія найближчих сусідів між $d$- і $f$-електронами та (vi) корельований перенос. Широко обговорюється відповідність числових результатів, отриманих для різних узагальнень моделі Фалікова-Кімбала, для опису реальних $d$-$f$ матеріалів.

\keywords модель Фалікова-Кімбалла, квантові конденсати, одновимірні
системи 
\end{abstract}
%\thanks{PACS nrs.: 71.27.+a, 71.28.+d, 71.30.+h}
\end{document}